\documentclass[]{aa}
\usepackage{graphics,latexsym,epsfig,graphicx,amssymb,natbib}     

\bibpunct{(}{)}{;}{a}{}{,}
\renewcommand{\d}[0]{{\rm d}}

\newcommand{\ave}[1]{\left\langle #1\right\rangle}

\newcommand{\Ref}[1]{(\ref{#1})}

\newcommand{\pprime}[0]{{\prime\prime}}
\newcommand{\snd}[0]{$2^{\rm nd}$-order~}
\newcommand{\thrd}[0]{$3^{\rm rd}$-order~}
\newcommand{\e}[0]{{\rm e}}
\newcommand{\msol}[0]{{\rm M}_\odot}
\renewcommand{\i}[0]{{\rm i}}
\newcommand{\changed}[1]{#1}
\begin{document}

\title{First detection of galaxy-galaxy-galaxy lensing in
  RCS\thanks{Based on observations from the Canada-France-Hawaii
  Telescope, which is operated by the National Research Council of
  Canada, le Centre Nationale de la Recherche Scientifique and the
  University of Hawaii.}} 

  \subtitle{A new tool for studying the matter environment of galaxy pairs}

\author{\mbox{P. Simon$^{1,2}$}, \mbox{P. Watts$^2$}, \mbox{P. Schneider$^2$},
  \mbox{H. Hoekstra$^3$}, \mbox{M.D. Gladders$^4$}, \mbox{H.K.C. Yee$^5$},
  \mbox{B.C. Hsieh$^6$}, \mbox{and H. Lin$^7$}}

\institute{%
  $^1$The Scottish Universities Physics Alliance (SUPA), Institute for
  Astronomy, School of Physics, University of Edinburgh, Royal
  Observatory, Blackford Hill, Edinburgh EH9 3HJ, UK\\
  $^2$  Argelander-Institut f\"ur
  Astronomie\thanks{Founded by merging of the Institut f\"ur
    Astrophysik und Extraterrestrische Forschung, the Sternwarte, and
    the Radioastronomisches Institut der Universit\"at Bonn.}, Universit\"at Bonn,
  Auf dem H\"ugel 71, 53121 Bonn, Germany\\
  $^3$ Department of Physics and Astronomy, University of Victoria,
  Victoria, BC, V8P 5C2, Canada\\
  $^4$ Department of Astronomy \& Astrophysics, University of Chicago,
  5640 S. Ellis Ave., Chicago, IL, 60637, US\\
  $^5$ Department of Astronomy \& Astrophysics, University of Toronto,
  60 St. George Street, Toronto, Ontario M5S 3H8, Canada\\
  $^6$ Institute of Astrophysics \& Astronomy, Academia Sinica,
  P.O. Box 23-141, Taipei 106, Taiwan, R.O.C.\\
  $^7$ Fermi National Accelerator Laboratory, P.O. Box 500, Batavia,
  IL 60510}

\date{Received/Accepted} \authorrunning{P. Simon} \titlerunning{}

\date{} 
\authorrunning{Simon et al.} 
\titlerunning{GGGL in RCS}


\abstract
{The weak gravitational lensing effect, small coherent distortions of
  galaxy images by means of a gravitational tidal field, can be used
  to study the relation between the matter and galaxy distribution.}
{In this context, weak lensing has so far only been used for
  considering a second-order correlation function that relates the
  matter density and galaxy number density as a function of
  separation. We implement two new, third-order correlation functions
  that have recently been suggested in the literature, and apply them
  to the Red-Sequence Cluster Survey. As a step towards exploiting
  these new correlators in the future, we demonstrate that it is
  possible, even with already existing data, to make significant
  measurements of third-order lensing correlations.}
{We develop an optimised computer code for the correlation
  functions. To test its reliability a set of tests involving mock
  shear catalogues are performed. The correlation functions are
  transformed to aperture statistics, which allow easy tests for
  remaining systematics in the data. In order to further verify the
  robustness of our measurement, the signal is shown to vanish when
  randomising the source ellipticities. Finally, the lensing signal
  is compared to crude predictions based on the halo-model.}
{On angular scales between \mbox{$\sim1^\prime$} and $\sim11^\prime$ a
  significant third-order correlation between two lens positions and
  one source ellipticity is found. We discuss this correlation
  function as a novel tool to study the average matter environment
  of pairs of galaxies. Correlating two source ellipticities and one
  lens position yields a less significant but nevertheless detectable
  signal on a scale of \mbox{$\sim4^\prime$}. Both signals lie roughly
  within the range expected by theory which supports their
  cosmological origin. }
{}

\maketitle
\section{Introduction}

The continuous deflection of light rays propagating through the
Universe by inhomogeneities of the large-scale distribution of matter
generates a coherent, weak distortion pattern over the sky: the shear
field. As the source of the shear field is gravity, the total matter
distribution contributes to this effect. It is, therefore, a probe of
the full matter content of the Universe. The shear field can be
investigated by correlating the shapes of distant faint galaxies
\citep[e.g.][]{sasfee01,waerbeke03, bs01}.  Since the first detection
of this effect \citep{2000MNRAS.318..625B, 2000astro.ph..3338K,
2000A&A...358...30V, 2000Natur.405..143W} studying the weak
gravitational lensing effect has become an important tool for
cosmology. Together with the next-generation weak lensing surveys such
as the \mbox{RCS2} (ongoing), \mbox{CFHTLS} (ongoing),
\mbox{Pan-STARRS 1} (commencing soon) or \mbox{KIDS} \changed{(planned
to commence during the second half of 2008)}, the weak lensing effect
will allow us to put tight constraints on cosmological parameters
\citep{eso06}.

One important topic in contemporary cosmology is the relation between
the dark matter and the galaxy population, the latter of which is
thought to form under particular conditions from the baryonic
component within the dark matter density field. This relation can be
studied by cross-correlating the shear signal and (angular) positions
of a selected galaxy population.

As the shear is quite a noisy observable, higher order galaxy-shear
correlation functions are increasingly difficult to measure. For this
reason, studies in the past have focused on \snd statistics
(``galaxy-galaxy lensing'', GGL hereafter) which involve one galaxy of
the selected population (foreground) and one source galaxy
(background) whose ellipticity carries the lensing signal. The
GGL-signal can be used to learn more about the typical dark matter
environment of single galaxies \citep[most
recently][]{klein06,mandel06a,mandel06b,mandel06c,
seljak05,2005ApJ...635...73H,2004ApJ...606...67H,sheldon04}, or the
so-called galaxy biasing \citep{simon07,pen03,hoekstra02,
2001ApJ...558L..11H}.

\citet{gggl05} introduced ``galaxy-galaxy-galaxy lensing'' (GGGL)
correlation functions and estimators thereof which allow us to move to
the next, \thrd level \citep[see also][]{watts05}. The correlation
functions now involve either two foreground galaxies and one
background galaxy, or one foreground galaxy and two background
galaxies. This idea was also discussed by \citet{2006MNRAS.367.1222J}
who studied how to derive the galaxy-galaxy-mass correlation function,
which is one of the foregoing two, from weak gravitational lensing.
These functions, although more difficult to measure than the two-point
GGL signal, offer the opportunity to study the typical environment of
pairs of galaxies, e.g., within galaxy groups (or more technically,
the occupation statistics of galaxies in dark matter halos,
\changed{see \citet{cooraysheth02} for a recent review}), or possibly
even the shape of dark matter haloes \citep{smith05}. More generally,
they measure \thrd moments between number densities of galaxies and
the matter density of dark matter (cross-correlation bispectra).
Hence, they ``see'' the lowest-order \emph{non-Gaussian} features
produced by cosmic structure formation.

This paper applies for the first time the GGGL-correlation functions
to existing data, the Red-Sequence Cluster Survey
\citep[RCS;][]{rcs05}, and demonstrates that with the current
generation weak lensing surveys it is already possible to extract
these particular \thrd statistics.

The outline of the paper is as follows.  We will give a brief
description of the survey in Sect.  \ref{datasection}. In
Sect. \ref{methodsection}, we will define the correlation functions
and their practical implementation as estimators for real data.  In
Sect. \ref{resultsection}, our results will be presented, discussed
and compared to halo-model based predictions to verify if the signal
has roughly the expected order of magnitude. Finally, in the same
section, we demonstrate how the GGGL correlation function involving
two lenses and one source can be used to map out the excess of matter
-- compared to the haloes of individual lenses -- about pairs of
lenses.

Wherever a specific fiducial cosmology is needed \mbox{$\Omega_{\rm
m}=0.3$}, for the matter density parameter, and
\mbox{$\Omega_\Lambda=0.7$}, for the dark energy \changed{density}
parameter, are assumed. \changed{Dark Energy is assumed to behave like
a cosmological constant.} For the dark matter power spectrum
normalisation we adopt $\sigma_8=0.9$.

\section{Data: The Red-Sequence Cluster Survey}
\label{datasection}

\begin{figure}
  \begin{center}
    \epsfig{file=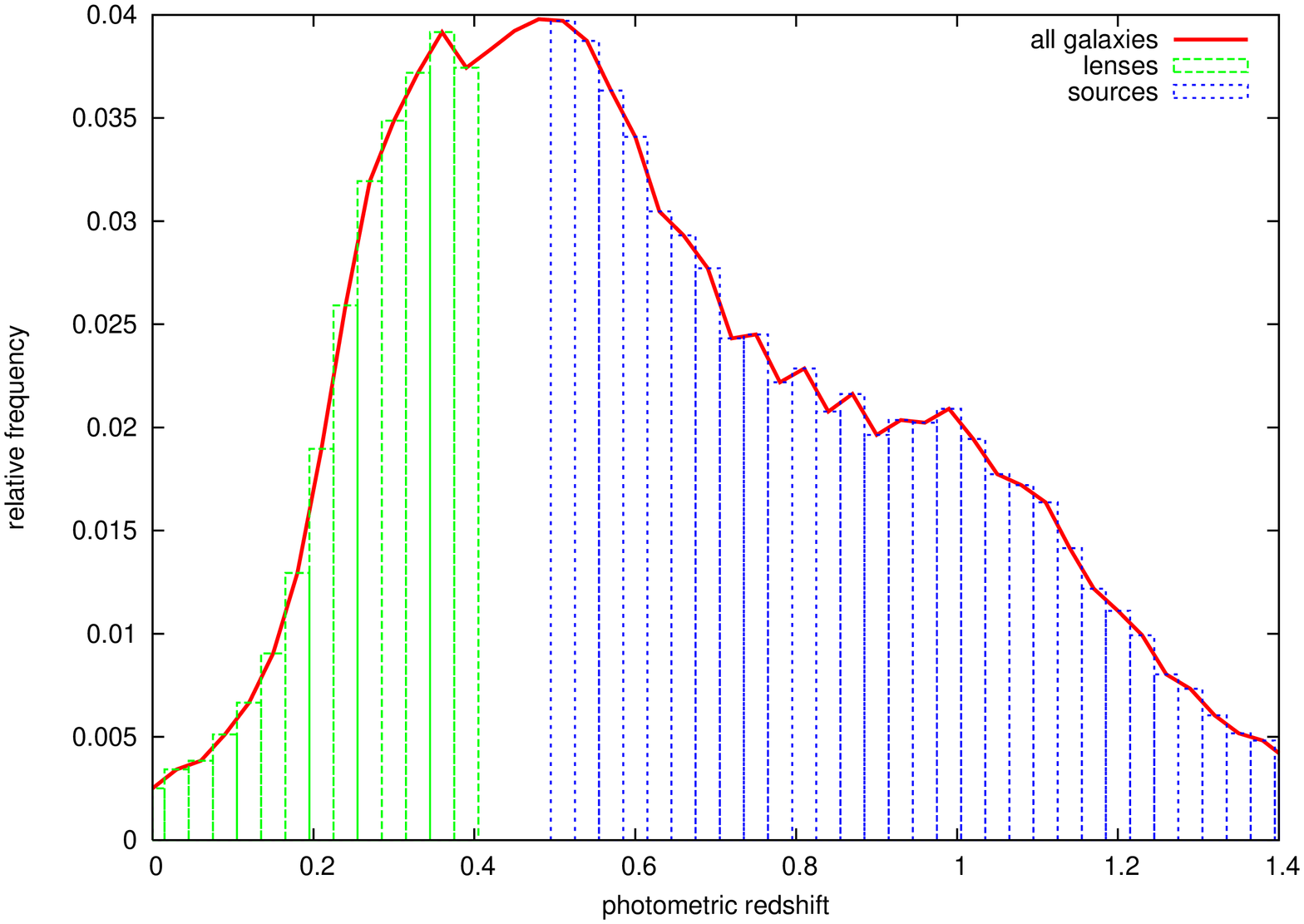,width=60mm,angle=-90}
  \end{center}
  \caption{\label{pofz}Histogram of photometric redshifts of lenses,
    \mbox{$z\in[0,0.4]$}, and sources, \mbox{$z\in[0.5,1.5]$}, used in
    our analysis. In total, we have got \mbox{$2.4\times10^5$} lenses
    (\mbox{$\bar{z}\approx0.30$}) and \mbox{$3.8\times10^5$} sources
    (\mbox{$\bar{z}\approx0.85$}).}
\end{figure}

The data used in this paper were taken as part of the Red-Sequence
Cluster Survey \citep[RCS;][]{rcs05}, and comprise of approximately
$34$ square \changed{degrees} of $B,V, R_C$ and $z'$ imaging data
observed with the Canada-France-Hawaii Telescope (CFHT). The $B$ and
$V$ bands were taken after completion of the original RCS, to allow
for a better selection of clusters at low redshifts. These follow-up
observations also enable the determination of photometric redshifts
for a large sample of galaxies. This photometric redshift information
is key for the work presented here. A detailed discussion of these
multicolour data, the reduction, and the photometric redshift
determination can be found in \citet{2005ApJS..158..161H}. In the
redshift range out to $z\sim 0.4$ the photometric redshifts are well
determined, with 70\% of the galaxies within 0.06 of the spectroscopic
redshift (as determined by comparing to a spectroscopic training set).
\changed{For fainter galaxies the uncertainties become naturally
larger. The photo-z uncertainty distribution in the RCS1 photo-z
catalogue is more or less a Gaussian for a given redshift range or a
given apparent magnitude range. The relation between photo-z
uncertainty, $\delta z$, and redshift is $\delta z\sim0.06(1+z)$. This
relation over-estimates the uncertainty for $z<0.7$ and
under-estimates it for $z>0.9$ since the systematic error gets larger
beyond that redshift.}

This photometric redshift catalogue was used by
\citet{2005ApJ...635...73H} to study the virial masses of isolated
galaxies as a function of luminosity and colour. To measure this
galaxy-galaxy lensing signal, the photometric redshift catalogue was
matched against the catalogue of galaxies for which shapes were
measured. This resulted in a sample of $8\times 10^5$ galaxies with
$18<R_c<24$ that are used in the analysis presented
here. \citet{2005ApJ...635...73H} \changed{also present} a number of
lensing specific tests, demonstrating the usefulness of the RCS
photometric redshift catalogue for galaxy-galaxy lensing studies. The
frequency distribution of photometric redshifts in our galaxy samples
is shown in Fig. \ref{pofz}.

The galaxy shapes were determined from the $R_C$ images. The raw
galaxy shapes are corrected for the effects of the
point-spread-function (PSF) as described in
\citet{1998ApJ...504..636H,2002ApJ...572...55H}. We refer the reader
to these papers for a detailed discussion of the weak lensing
analysis. We note that the resulting object catalogues have been used
for a range of weak lensing studies. Of these, the measurements of the
lensing signal caused by large scale structure presented in
\citet{2002ApJ...577..595H,2002ApJ...572...55H} are particularly
sensitive to residual systematics. The various tests described in
these papers suggest that the systematics are well under control. It
is therefore safe to conclude that residual systematics in the galaxy
shape measurements are not a significant source of error in the
analysis presented here.

\section{Method}
\label{methodsection}

Here we briefly summarise definitions of the three-point correlation
functions, their estimators and the relation between aperture
statistics and correlation functions. A detailed derivation and
explanation of those can be looked up in \citet{gggl05}.

\subsection{GGGL-correlation functions}

\begin{figure*}
  \begin{center}
    \epsfig{file=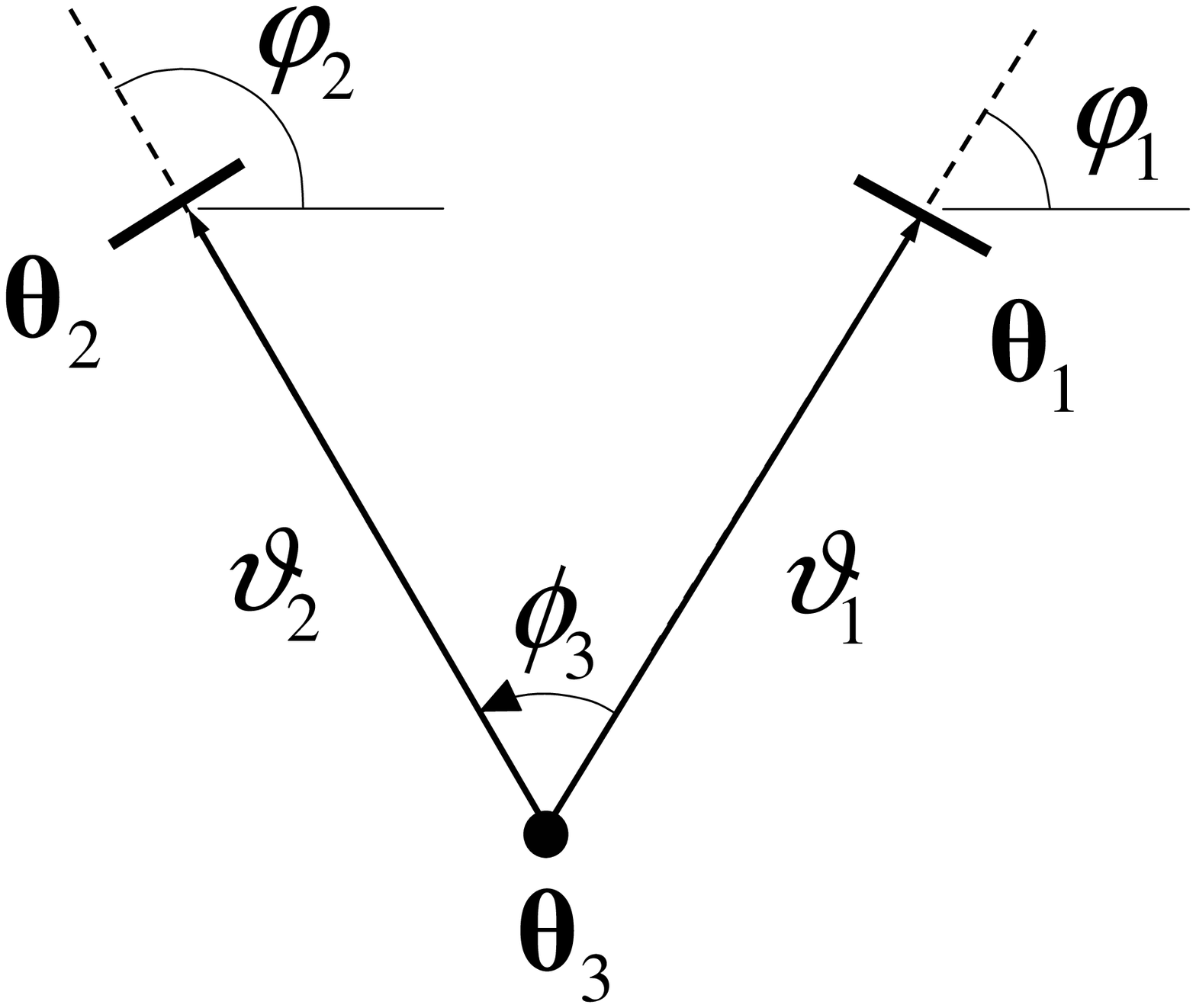,width=65mm,angle=0}
    \hspace{2cm}
    \epsfig{file=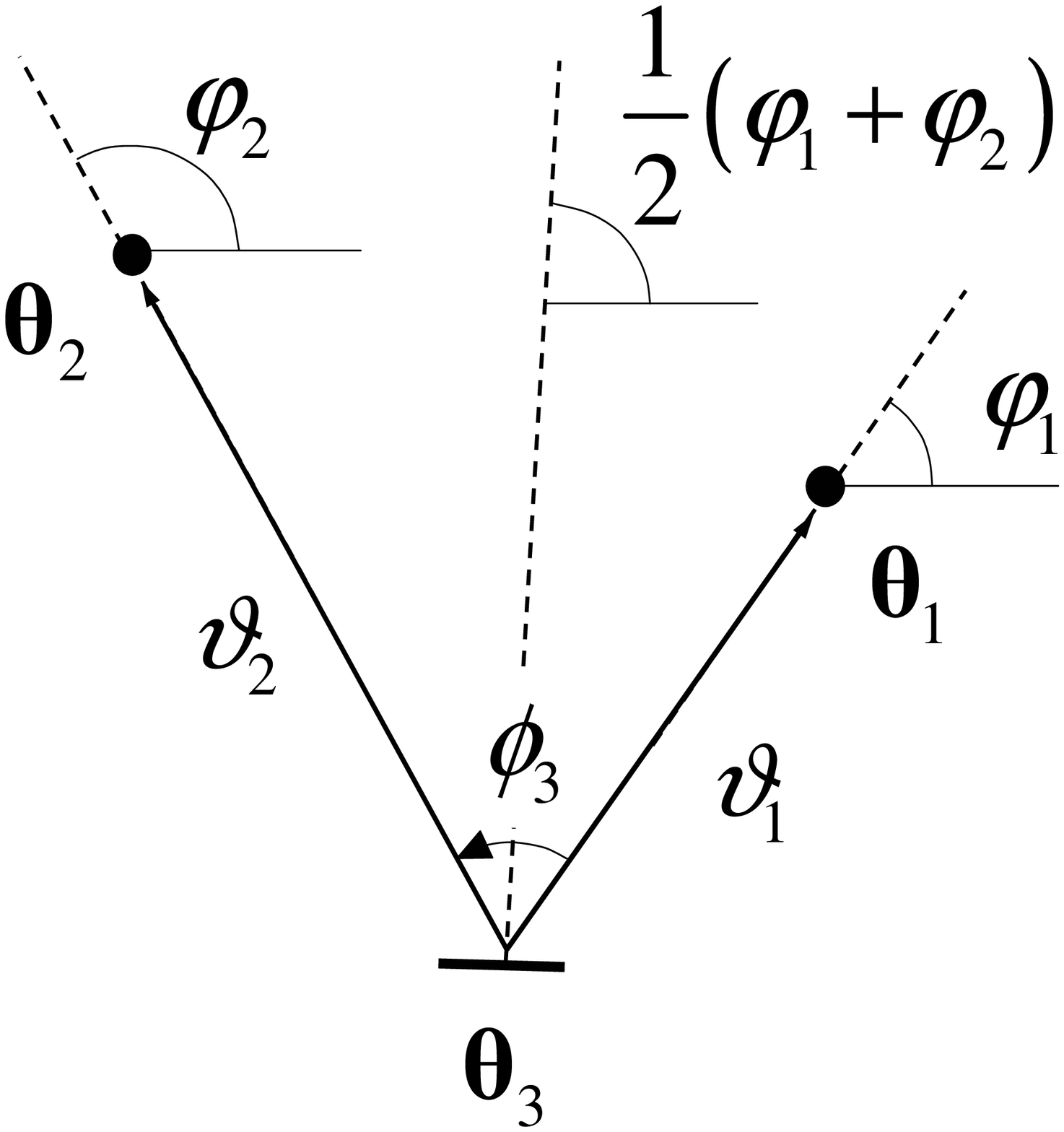,width=65mm,angle=0}
  \end{center}
  \caption{\label{corrsketch}Geometry of the galaxy-shear-shear
    correlation, $G_\pm(\vartheta_1,\vartheta_2,\phi_3)$ (\emph{left
      panel}), and the galaxy-galaxy-shear correlation, ${\cal
      G}(\vartheta_1,\vartheta_2,\phi_3)$ (\emph{right panel}). The
    figure is copied from \citet{gggl05}.}
 \end{figure*}

 For our analysis we consider two different classes of correlation
 functions. Both classes require triplets of galaxies which are
 located at the positions $\vec{\theta}_1$, $\vec{\theta}_2$ and
 $\vec{\theta}_3$ on the sky (see Fig. \ref{corrsketch}). In a
 cosmological context, random fields -- such as the projected number
 density of galaxies, $N(\vec{\theta})$, or the shear field,
 $\gamma(\vec{\theta})$ -- are statistically homogeneous and
 isotropic. For that reason, all conceivable correlations between the
 values of those fields depend merely on the separation,
 \mbox{$|\vec{\theta}_i-\vec{\theta}_j|$}, and never on absolute
 positions $\vec{\theta}_i$. Therefore, our correlators are solely
 functions of the dimensions of the triangle formed by the
 galaxies. We parameterise the dimension of a triangle in terms of the
 lengths of two triangle edges, $\vartheta_1$ and $\vartheta_2$, and
 one angle, $\phi_3$, that is subtended by the edges.  Note that the
 sign of $\phi_3$, i.e. the \changed{handedness} of the triangle, is
 important.

 The \emph{galaxy-galaxy-shear correlator}, 
 \begin{equation}\label{g}
   {\cal G}(\vartheta_1,\vartheta_2,\phi_3)=
   \ave{
     \kappa_{\rm g}(\vec{\theta}_1)\kappa_{\rm g}(\vec{\theta}_2)
     \gamma\left(\vec{\theta}_3;\frac{\varphi_1+\varphi_2}{2}\right)}\;,      
 \end{equation}
 is the expectation value of the shear at $\vec{\theta}_3$ rotated in
 the direction of the line bisecting the angle $\phi_3$ multiplied by
 the number density contrast of lens (foreground) galaxies at
 $\vec{\theta}_{1,2}$:
 \begin{equation}
   \kappa_{\rm g}(\vec{\theta})\equiv\frac{N(\vec{\theta})}{\overline{N}}-1\;.
 \end{equation}
 A rotation of shear is defined as
 \begin{equation}
   \gamma\left(\vec{\theta};\varphi\right)\equiv
   -\e^{-2\i\varphi}\gamma_{\rm c}(\vec{\theta})\;,
 \end{equation}
 where $\gamma_{\rm c}$ is the shear relative to a Cartesian coordinate
 frame.  It should be noted that $\cal G$ and the following
 correlators are complex numbers.

 A second class of correlators are the \emph{galaxy-shear-shear
   correlators},
 \begin{eqnarray}\nonumber
   G_+(\vartheta_1,\vartheta_2,\phi_3)&=&
   \ave{
     \gamma(\vec{\theta}_1;\varphi_1)
     \gamma^\ast(\vec{\theta}_2;\varphi_2)
     \kappa_{\rm g}(\vec{\theta}_3)}\;,\\\label{gpm}
   G_-(\vartheta_1,\vartheta_2,\phi_3)&=&
   \ave{
     \gamma(\vec{\theta}_1;\varphi_1)
     \gamma(\vec{\theta}_2;\varphi_2)
     \kappa_{\rm g}(\vec{\theta}_3)}\;,   
 \end{eqnarray}
 which correlate the shear at two points with the lens galaxy number
 density contrast at another point. Again, the shears are rotated,
 this time in the direction of the lines connecting the source
 (background) galaxies, at $\vec{\theta}_{1,2}$, and the lens galaxy
 at $\vec{\theta}_3$.

\subsection{Practical estimators of correlators}
\label{practise}

With practical estimators for \Ref{g} and \Ref{gpm} in mind,
\citet{gggl05} introduced modified correlation functions. They differ
from $\cal G$ and $G_\pm$ in that they are defined in terms of the
number density of the lens galaxies, $N(\vec{\theta})$, instead of the
number density contrast, $\kappa_{\rm g}$:
\begin{eqnarray}\label{gtilde}
  &&\tilde{\cal G}(\vartheta_1,\vartheta_2,\phi_3)\equiv
  \frac{\ave{N(\vec{\theta}_1)N(\vec{\theta}_2)
    \gamma\left(\vec{\theta}_3;\frac{\varphi_1+\varphi_2}{2}\right)}}
  {\overline{N}^2}
  \\\nonumber
  &&={\cal G}(\vartheta_1,\vartheta_2,\phi_3)+
  \ave{\gamma_{\rm t}}(\vartheta_1)\e^{-\i\phi_3}+
  \ave{\gamma_{\rm t}}(\vartheta_2)\e^{+\i\phi_3}\;,
\end{eqnarray}
 and
\begin{eqnarray}
  &&\tilde{G}_+(\vartheta_1,\vartheta_2,\phi_3)\equiv
  \frac{1}{\overline{N}}
  \ave{
    \gamma(\vec{\theta}_1;\varphi_1)
    \gamma^\ast(\vec{\theta}_2;\varphi_2)
    N(\vec{\theta}_3)}\\\nonumber
  &&=G_+(\vartheta_1,\vartheta_2,\phi_3)+
  \ave{\gamma(\vec{\theta}_1;\varphi_1)\gamma^\ast(\vec{\theta}_2;\varphi_2)}\;,\\
  \nonumber\\
  &&\tilde{G}_-(\vartheta_1,\vartheta_2,\phi_3)\equiv
  \frac{1}{\overline{N}}
  \ave{
    \gamma(\vec{\theta}_1;\varphi_1)
    \gamma(\vec{\theta}_2;\varphi_2)
    N(\vec{\theta}_3)}\\\nonumber
  &&=G_-(\vartheta_1,\vartheta_2,\phi_3)+
  \ave{\gamma(\vec{\theta}_1;\varphi_1)\gamma(\vec{\theta}_2;\varphi_2)}\;.
\end{eqnarray}
These correlators \changed{also} contain, apart from the original
purely \thrd contributions, contributions from \snd correlations:
$\ave{\gamma_{\rm t}}(\theta)$ is the mean tangential shear about a
single lens galaxy at separation $\theta$ (GGL),
$\ave{\gamma(\vec{\theta}_1;\varphi_1)
\gamma^\ast(\vec{\theta}_2;\varphi_2)}$ and
$\ave{\gamma(\vec{\theta}_1;\varphi_1)
\gamma(\vec{\theta}_2;\varphi_2)}$ are shear-shear correlations which
are functions of the cosmic-shear correlators $\xi_\pm(\theta)$
\citep[e.g.][]{bs01}.  To recover pure \thrd statistics, the \snd
terms can either be subtracted, or even neglected, if we work in terms
of the aperture statistics, as we will see in the next section.

With respect to practical estimators, number densities are more useful
quantities because every single galaxy position is an unbiased
estimator of $N(\vec{\theta})/\overline{N}$. For that reason, every
triangle of galaxies that can be found in a survey can be made an
unbiased estimator of either $\tilde{\cal G}$ (two lenses and one
source) or $\tilde{G}_\pm$ (two sources and one lens). Since,
generally, a weighted average of (unbiased) estimates is still an
(unbiased) estimate\footnote{The weighting scheme only influences the
statistical uncertainty of the average, i.e. the variance of the
combined estimate. Note that the whole statement requires that the
weights are uncorrelated with the estimates that the average is taken
of.}, we can combine the estimates of all triangles of the same
dimension using arbitrary weights, $w_{j/k}$, for the sources. Note
that for the following sums only triangles of the same $\vartheta_1$,
$\vartheta_2$ and $\phi_3$ have to be taken into account inside the
sums. We adopt a binning such that $\vartheta_1$, $\vartheta_2$ and
$\phi_3$ need to be within some binning interval to be included inside
the sums, i.e.  triangles of similar dimensions are used for the
averaging:
\begin{eqnarray}\label{gpestimator}
  \tilde{G}^{\rm est}_+&=&
  \frac{
    \sum\limits_{i,j,k=1}^{N_{\rm l},N_{\rm s}}
    w_j\,w_k\,\epsilon_j\epsilon_k^\ast\e^{-2\i\varphi_j}\e^{+2\i\varphi_k}}{
  \sum\limits_{i,j,k=1}^{N_{\rm l},N_{\rm s}}w_j\,w_k}\;,\\
\label{gmestimator}
  \tilde{G}^{\rm est}_-&=&
  \frac{
    \sum\limits_{i,j,k=1}^{N_{\rm l},N_{\rm s}}
    w_j\,w_k\,\epsilon_j\epsilon_k\e^{-2\i\varphi_j}\e^{-2\i\varphi_k}}{
  \sum\limits_{i,j,k=1}^{N_{\rm l},N_{\rm s}}w_j\,w_k}\;,
\end{eqnarray}
where $j,k\in\{1\ldots N_{\rm s}\}$ are indices for sources and
    $i\in\{1\ldots N_{\rm l}\}$ is the index of the lenses; $N_{\rm
    l}$ and $N_{\rm s}$ are the number of lenses and sources,
    respectively. By $\varphi_j$ and $\varphi_k$ we denote the phase
    angles of the two sources relative to the foreground galaxy $i$.

The statistical weights are chosen to down-weight triangles that
contain sources whose complex ellipticities, $\epsilon_i$
\citep{bs01}, are only poorly determined. Lenses, however, always have
the same weight in our analysis.

Similarly, we can define an estimator for $\tilde{\cal G}$.  However,
one has to take into account that
\changed{\mbox{$\epsilon\,\e^{-\i(\varphi_1+\varphi_2)}$} of one
single triangle} -- consisting of two lenses and one source with
ellipticity $\epsilon$ -- is an estimator of
\begin{equation}\label{rescale}
  \frac{\ave{
      N(\vec{\theta}_1)N(\vec{\theta}_2)
      \gamma\left(\vec{\theta}_3;\frac{\varphi_1+\varphi_2}{2}\right)}}
  {\ave{N(\vec{\theta}_1)N(\vec{\theta}_2)}}=
  \frac{\tilde{\cal G}(\vartheta_1,\vartheta_2,\phi_3)}
  {1+\omega(|\vec{\theta}_2-\vec{\theta}_1|)}  
\end{equation}
and \emph{not} $\tilde{\cal G}$ alone as has falsely been assumed in
\citet{gggl05}.\footnote{This becomes apparent if one sets
\mbox{$\gamma\left(\vec{\theta}_i;\frac{\varphi_1+\varphi_2}{2}\right)=
\gamma\left(\vec{\theta}_3;\frac{\varphi_1+\varphi_2}{2}\right)=\rm
const$} in the Eqs.  (34) and (32), respectively, of \citet{gggl05}.}
The function
\begin{equation}
  \omega(|\vec{\Delta\theta}|)\equiv
  \ave{\kappa_{\rm g}(\vec{\theta})\kappa_{\rm g}(\vec{\theta}+\vec{\Delta\theta})}
\end{equation}
is the angular clustering of the lenses \citep{peebles80}.  Based on
this notion, we can write down an estimator for $\tilde{\cal G}$:
\begin{equation}\label{gestimator}
  \tilde{\cal G}^{\rm est}=
  \frac{\sum\limits_{i,j,k=1}^{N_{\rm l},N_{\rm s}}
    w_k\,
    \epsilon_k\,\e^{-\i(\varphi_i+\varphi_j)}
    \left[1+\omega(|\vec{\theta}_i-\vec{\theta}_j|)\right]}
  {(-1)\,\sum\limits_{i,j,k=1}^{N_{\rm l},N_{\rm s}}w_k}
\end{equation}
that includes explicitly the clustering of lenses.  Here, $w_k$
($k\in\{1\ldots N_{\rm s}\}$) are the statistical weights of the
sources. By $\varphi_i$ and $\varphi_j$ ($i,j\in\{1\ldots N_{\rm
l}\}$) we denote the phase angles of the two lenses relative to the
source $k$. Again, only triangles of the same or similar dimensions
(\changed{parameters in same bins}) are to be included inside the
sums.

\begin{figure*}
  \begin{center}
    \includegraphics[width=14cm, angle=0]{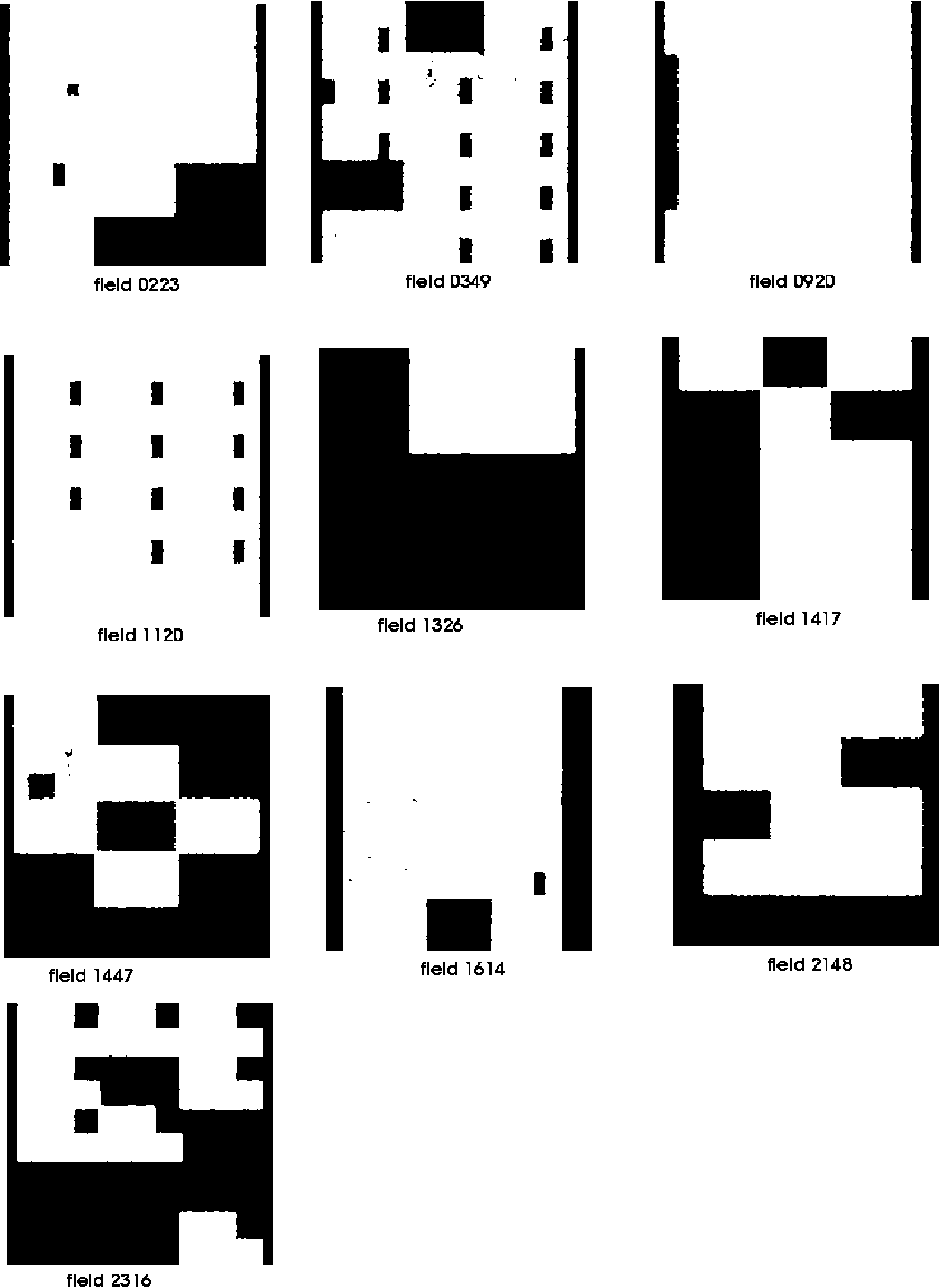}
  \end{center}
  \caption{\label{masks}Determined geometry of the individual RCS
    fields (field masks). Each mask has a size of approximately
    \mbox{$139^\prime\times139^\prime$} or, equivalently,
    \mbox{$20.000\times20.000~\rm pixel^2$}. Black colours denote
    \changed{masked-out} regions, while white indicates regions where
    galaxies have been found.}
\end{figure*}

\changed{For obtaining an estimate of $\omega(\theta)$ in practice we
  employed the estimator of \citet{ls93}, which, compared to other
  estimators, minimises the variance to nearly Poissonian}:
\begin{equation}
  \label{omega}
  \omega\left(\theta\right)=\frac{DD}{RR}-2\frac{DR}{RR}+1\;.
\end{equation}
It requires \changed{one} to count the number of (lens) galaxy pairs
with a separation between $\theta$ and \mbox{$\theta+\delta\theta$},
namely the number of pairs in the data, denoted by $DD$, the number of
pairs in a random mock catalogue, $RR$, and the number of pairs that
can be formed with one data galaxy and one mock data galaxy, $DR$. The
random mock catalogue is computed by randomly placing the galaxies,
taking into account the geometry of the data field, i.e. by avoiding
masked-out regions, see Fig. \ref{masks}. We generate $25$ random
galaxy catalogues and average the pair counts obtained for $DR$ and
$RR$.

When computing the $\tilde{\cal G}$ and $\tilde{G}_\pm$ estimators, we
suggest the use of complex numbers for the angular positions of
galaxies: \mbox{$\vec{\vartheta}=\vartheta_1+\i\vartheta_2$} with
$\vartheta_{1,2}$ being the $x$/$y$-coordinates relative to some
Cartesian reference frame (flat-sky approximation). The phase factors
turning up inside the sums \Ref{gpestimator}, \Ref{gmestimator} and
\Ref{gestimator} are then simply (notation of Fig. \ref{corrsketch}):
\begin{equation}
  \e^{-2\i\varphi_1}=\frac{\vec{\vartheta}^\ast_{13}}{\vec{\vartheta}_{13}}\,;\,
  \e^{-2\i\varphi_2}=\frac{\vec{\vartheta}^\ast_{23}}{\vec{\vartheta}_{23}}\,;\,
  \e^{-\i(\varphi_1+\varphi_2)}=
  \frac{\vec{\vartheta}_{13}\vec{\vartheta}_{23}}
  {|\vec{\vartheta}_{13}||\vec{\vartheta}_{23}|}\;,
\end{equation}
where \mbox{$\vec{\vartheta}_{ij}\equiv\vec{\vartheta_i}-\vec{\vartheta}_j$}.

\subsection{Conversion to aperture statistics}
\label{apsection}

In weak lensing, cosmological large-scale structure is often studied
in terms of the aperture statistics
\citep{simon07,kill05,jarvis04,hoekstra02,schneider98,ludo98} that
measure the convergence (projected matter distribution), $\kappa$, and
projected number density fields of galaxies, $\kappa_{\rm g}$,
\changed{smoothed with a compensated filter} $u(x)$, i.e. $\int_0^\infty\d x\,xu(x)=0$:
\begin{eqnarray}\label{map}
  M_{\rm ap}(\theta)&=&
  \frac{1}{\theta^2}
  \int_0^\infty\,\d^2\vartheta\,u\!\left(\frac{|\vec{\vartheta}|}{\theta}\right)\,\kappa(|\vec{\vartheta}|)\;,\\\label{nap}
  {\cal N}(\theta)&=&
  \frac{1}{\theta^2}
  \int_0^\infty\,\d^2\vartheta\,u\!\left(\frac{|\vec{\vartheta}|}{\theta}\right)\,\kappa_{\rm g}(|\vec{\vartheta}|)\;,
\end{eqnarray}
where $\theta$ is the smoothing radius. $M_{\rm ap}$ is called the
aperture mass, while ${\cal N}$ is the aperture number count of
galaxies. With an appropriate filter these aperture measures are only
sensitive to a very narrow range of spatial Fourier modes so that they
are extremely suitable for studying the scale-dependence of structure,
or even the scale-dependence of remaining systematics in the data
\citep{hetter06}. Moreover, they provide a very localised measurement
of power spectra (band power), in the case of $\ave{{\cal N}^nM_{\rm
ap}^m}$ for $n+m=2$, and bispectra, in the case of $n+m=3$, without
relying on complicated transformations between correlation functions
and power spectra. The aperture filter we employ for this paper is:
\begin{equation}\label{filter}
  u(x)=\frac{1}{2\pi}\left(1-\frac{x^2}{2}\right)\e^{-x^2/2}
\end{equation}
as introduced by \citet{critt02}. For an aperture radius of $\theta$
the filter peaks at a spatial \changed{wavelength} of
\mbox{$\ell=\frac{\sqrt{2}}{\theta}$} which corresponds to a typical
angular scale of
\mbox{$\frac{2\pi}{\ell}=\frac{\pi}{\sqrt{2}}\theta$}.

As shear and convergence are \changed{both linear combinations of
second derivatives of the deflection potential}, the aperture mass can
be computed from the shear in the following manner
\citep{1998MNRAS.296..873S}:
\begin{eqnarray}\label{mapinshear}
  M_{\rm ap}(\theta)+\i M_\perp(\theta)&=&
  \frac{1}{\theta^2}\int_0^\infty\!\!\!\!\d^2\vec{\theta}^\prime\,
  q\left(\frac{|\vec{\theta}^\prime|}{\theta}\right)
  \gamma\left(\vec{\theta}^ \prime;\angle\vec{\theta}^\prime\right)\!\!\;,\\
  q(x)&\equiv&\frac{2}{x^2}\int_0^x\d s\,s\,u(s)-u(x)\;,
\end{eqnarray}
where we denote by $\angle\vec{\theta}^\prime$ the polar angle of the
vector $\vec{\theta}^\prime$. Note that in Eq. \Ref{mapinshear} we
place, for convenience, the origin of the coordinate system at the
centre of the aperture.

In expression \Ref{mapinshear}, $M_{\rm ap}$ is the E-mode, whereas
$M_\perp$ is the B-mode of the aperture mass. Of central importance
for our work is that we can extract E- and B-modes of the aperture
statistics from the correlation functions.  Since B-modes cannot be
generated by weak gravitational lensing, a zero or small B-mode is an
important check for a successful PSF-correction of real data
\citep[e.g.][]{hetter06}, or the violation of parity-invariance in the
data \citep{schneider03}, which is also a signature of systematics.

Another argument in favour of using aperture statistics at this stage
of our analysis is that \snd terms in $\tilde{\cal G}$ and
$\tilde{G}_\pm$ do not contribute to the \thrd aperture statistics
\citep{gggl05}.  Therefore, a significant signal in the aperture
statistics means a true detection of \thrd correlations.

The \thrd aperture statistics can be computed from $\tilde{\cal G}$
via:
\begin{eqnarray}\label{n2mapemode}
 && \ave{{\cal N}^2M_{\rm ap}}(\theta_1,\theta_2,\theta_3)=\\\nonumber
 && \Re{\left({\cal I}\left[
  \tilde{\cal G}(\vartheta_1,\vartheta_2,\phi_3)
  A_{{\cal N}{\cal
      N}M}(\vartheta_1,\vartheta_2,\phi_3|\theta_1,\theta_2,\theta_3)
  \right]\right)}\;,\\\nonumber\\
\label{n2mappmode}
 && \ave{{\cal N}^2M_\perp}(\theta_1,\theta_2,\theta_3)=\\\nonumber
 && \Im{\left({\cal I}\left[
  \tilde{\cal G}(\vartheta_1,\vartheta_2,\phi_3)
  A_{{\cal N}{\cal
      N}M}(\vartheta_1,\vartheta_2,\phi_3|\theta_1,\theta_2,\theta_3)
  \right]\right)}\;,
\end{eqnarray}
where we have introduced for the sake of brevity an abbreviation for
the following integral:
\begin{equation}\label{integral}
  {\cal I}\left[f\right]\equiv
  \int\limits_0^\infty\d\vartheta_1\vartheta_1
  \int\limits_0^\infty\d\vartheta_2\vartheta_2
  \int\limits_0^{2\pi}\d\phi_3\,\,f\;.
\end{equation}
By $\Re{(x)}$ and $\Im{(x)}$ we denote the real and imaginary part,
respectively, of a complex number $x$.  Eq. \Ref{n2mapemode} is the
E-mode of the aperture moment $\ave{{\cal N}(\theta_1){\cal
N}(\theta_2)M_{\rm ap}(\theta_3)}$, whereas Eq.  \Ref{n2mappmode} is
the corresponding parity mode that is non-zero in the case of
violation of parity-invariance; the latter has to be zero \emph{even
if} B-modes are present in the shear pattern that may be produced to
some degree by intrinsic source alignment \citep[e.g.][]{heymans04} or
intrinsic ellipticity/shear correlations \citep{hirata04} -- that is,
if we assume that the macroscopic world is parity-invariant. The
integral kernel $A_{{\cal N}{\cal N}M}$ for our aperture filter can be
found in the Appendix of \citet{gggl05}.

The aperture statistics associated with the GGGL-correlator
$\tilde{G}_\pm$ are the following:
\begin{eqnarray}\label{nmap2emode}
  &&  \ave{M_{\rm ap}^2{\cal
      N}}(\theta_1,\theta_2,\theta_3)=\\\nonumber
  &&\Re{\left[
      \ave{MM{\cal N}}(\theta_1,\theta_2,\theta_3)+
      \ave{MM^\ast{\cal N}}(\theta_1,\theta_2,\theta_3)\right]}/2\;,\\\nonumber\\
  \label{nmap2bmode}
  &&  \ave{M_\perp^2{\cal
      N}}(\theta_1,\theta_2,\theta_3)=\\\nonumber
  &&\Re{\left[
      \ave{MM^\ast{\cal N}}(\theta_1,\theta_2,\theta_3)-
      \ave{MM{\cal N}}(\theta_1,\theta_2,\theta_3)\right]}/2\;,\\\nonumber\\
  \label{nmap2pmode}
  &&  \ave{M_\perp M_{\rm ap}{\cal
      N}}(\theta_1,\theta_2,\theta_3)=\\\nonumber
  &&\Im{\left[
      \ave{MM{\cal N}}(\theta_1,\theta_2,\theta_3)+
      \ave{MM^\ast{\cal N}}(\theta_1,\theta_2,\theta_3)\right]}/2\;,
\end{eqnarray}
where we used the following definitions
\begin{eqnarray}
  &&\ave{MM{\cal N}}(\theta_1,\theta_2,\theta_3)\equiv\\\nonumber
  &&{\cal I}\left[
    \tilde{G}_-(\vartheta_1,\vartheta_2,\phi_3)
    A_{MM{\cal
        N}}(\vartheta_1,\vartheta_2,\phi_3|\theta_1,\theta_2,\theta_3)\right]\;,\\\nonumber\\
  &&\ave{MM^\ast{\cal N}}(\theta_1,\theta_2,\theta_3)\equiv\\\nonumber
  &&{\cal I}\left[
    \tilde{G}_+(\vartheta_1,\vartheta_2,\phi_3)
    A_{MM^\ast{\cal
        N}}(\vartheta_1,\vartheta_2,\phi_3|\theta_1,\theta_2,\theta_3)\right]\;.
\end{eqnarray}
Eq. \Ref{nmap2emode} is the E-mode of $\ave{M_{\rm ap}(\theta_1)M_{\rm
    ap}(\theta_2){\cal N}(\theta_3)}$, Eq. \Ref{nmap2bmode} is the
    B-mode which should vanish if the shear pattern is purely
    gravitational, and Eq. \Ref{nmap2pmode} is again a parity-mode
    which is a unique indicator for systematics. As before, the
    integral kernels $A_{MM{\cal N}}$ and $A_{MM^\ast{\cal N}}$ for
    our aperture filter may be found in the Appendix of
    \citet{gggl05}.

\subsection{Validating the code}

\begin{figure}
  \begin{center}
    \includegraphics[width=65mm, angle=-90]{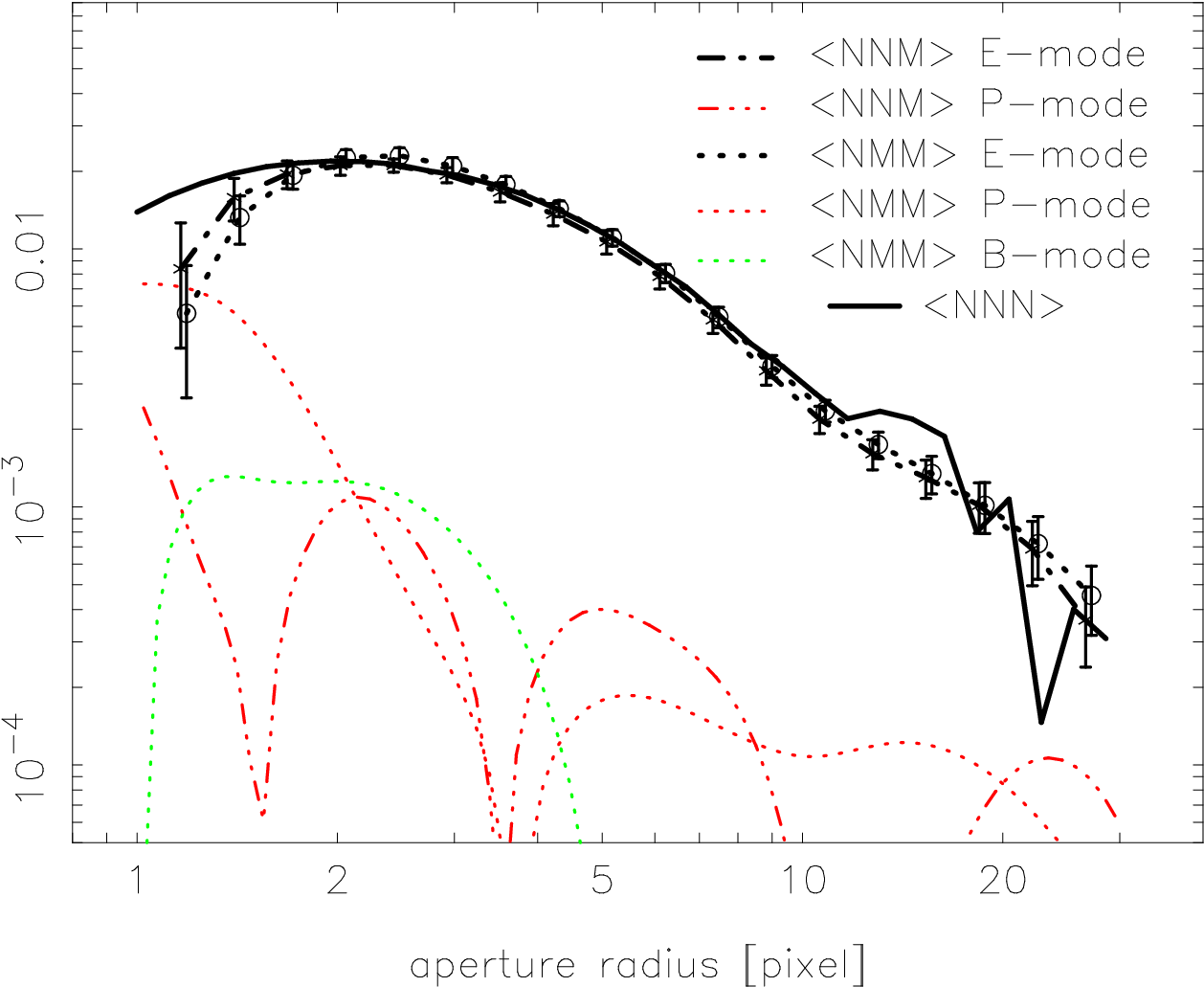}
  \end{center}
  \caption{\label{kappatest}Test run of our computer code with mock
    data based on some arbitrary convergence field. The mock data has
    been prepared \changed{such that} $\ave{{\cal N}^2(\theta)M_{\rm
    ap}(\theta)}= \ave{{\cal N}(\theta)M^2_{\rm
    ap}(\theta)}=\ave{{\cal N}^3(\theta)}$;
    \mbox{$\ave{NNN}\equiv\ave{{\cal N}^3(\theta)}$} is the value that
    has to be found by the code (only equally sized apertures are
    correlated for test: \mbox{$\theta_1=\theta_2=\theta_3=\theta$}).
    The binning range is between \mbox{$\vartheta\in[0.05,200]~\rm
    pixel$} with $100$ bins; we use $10^4$ lenses and the same number
    of sources.  For radii greater than $\sim\!2$ pixel we get good
    agreement. The expected signal (solid line; computed from placing
    apertures) becomes inaccurate beyond \mbox{$\theta\gtrsim10$
    pixel} because the aperture size becomes comparable to the field
    size. The error bars denote the \mbox{$1\sigma$} sampling
    uncertainty due to finite galaxy numbers. The B- and parity modes
    (P) of the statistics are two orders of magnitude smaller than the
    E-modes and are oscillating about zero (plotted is modulus).}
\end{figure}

In the last section, we outlined the steps which have to be undertaken
in order to estimate the \thrd aperture moments from a given catalogue
of lenses and sources. The three steps are: 1) estimating the angular
clustering of lenses yielding $\omega(\theta)$, 2) estimating
$\tilde{\cal G}$ and $\tilde{G}_\pm$ for some range of
$\vartheta_{1,2}$ and \changed{for $\phi_3\in[0,2\pi[$}, and finally
3) transforming the correlation function to $\ave{{\cal N}^2M_{\rm
ap}}$ and $\ave{{\cal N}M^2_{\rm ap}}$ including all E-, B- and
parity-modes.

There are several practical issues involved here. One issue is that,
in theory, for the transformation we require $\tilde{\cal G}$,
$\tilde{G}_\pm$ for all $\vartheta\in[0,\infty]$, see Eq.
\Ref{integral}. In reality, we will have both a lower limit (seeing,
galaxy-galaxy overlapping), $\vartheta_{\rm low}$, and an upper limit
(finite fields), $\vartheta_{\rm upper}$. On the other hand, the
GGGL-correlators drop off quickly for large $\vartheta$ and the
integral kernels $A_{{\cal N}{\cal N}M}$, $A_{MM{\cal N}}$,
$A_{MM^\ast{\cal N}}$ have exponential cut-offs for
$\vartheta_1,\vartheta_2\gg\theta_{1,2,3}$. Therefore, we can assume
that there will be some range where we can compute the aperture
statistics with \changed{satisfactory accuracy}. We perform the
following test to verify that this is true: by using theoretical
3D-bispectra of the galaxy-dark matter cross-correlations
\citep{watts05} we compute both the GGGL-correlation functions and the
corresponding aperture statistics \citep[Eqs. 37, 38, 40, 51, 52
of][]{gggl05}. By binning the GGGL-correlators we perform the
transformation including binning and cut-offs in $\vartheta$. We find
that one can obtain an accurate estimate of the aperture statistics
within a few percent between roughly
\mbox{$\theta\gtrsim40\,\vartheta_{\rm low}$} and
\mbox{$\theta\lesssim\vartheta_{\rm upper}/10$} (using $100$ log-bins
for $\vartheta_{1,2}$ and $100$ linear bins for $\phi_3$).  Therefore,
with RCS-fields of typical size $139^\prime$ we can expect to get an
accurate result between about
$0^\prime\!.5\lesssim\theta\lesssim14^\prime$.

Another issue is with \changed{step two above}, in which the
GGGL-correlators themselves need to be estimated. The estimators --
Eqs. \Ref{gpestimator}, \Ref{gmestimator} and \Ref{gestimator} -- in
terms of galaxy positions and source ellipticities are simple but the
enormous number of triangles that need to be considered is
computationally challenging (roughly $10^{13}$ per field for RCS). To
optimise this process we employ a data structure based on a binary
tree, a so-called tree code \citep[e.g.][]{jarvis04,zhang03}. The
tree-code represents groups of galaxies within some distance to a
particular triangle vertex as ``single galaxies'' with appropriate
weight (and average ellipticity).  This strategy effectively reduces
the number of triangles. Moreover, we optimise the code such that only
distinct triangles are found.  Then, the other triangle obtained be
exchanging the indices of either the two lenses ($\tilde{\cal G}$) or
the two sources ($\tilde{G}_\pm$) is automatically accounted for; this
reduces the computation time by a factor of two.

In order to test the performance and reliability of the code, we
create a catalogue of mock data. In order to do this we use a
simulated convergence field ($\kappa$-field) on a grid,
$512\times512~\rm pixel^2$, which has been obtained by ray-tracing
through an N-body simulated universe. Actually, the only requirement
that has to be met by the test field is that it behaves like a density
contrast $\delta$, i.e. \mbox{$\ave{\delta}=0$} and
\mbox{$\delta\geq-1$}, and that it has non-vanishing \thrd moments,
\mbox{$\ave{\delta^3}\ne0$}. Based on this field we simulate a shear
and lens catalogue. The shear catalogue is generated by converting the
$\kappa$-field to a shear field and by randomly selecting positions
within the field to be used as source positions. The positions and
associated shear provide the mock shear catalogue; for details see
\citet{simon04}. In a second step, we use the $\kappa$-field as
density contrast, $\kappa_{\rm g}$, of the number density of lenses to
make realisations of lens catalogues.  This means one randomly draws
positions, $\vec{\theta}$, within the grid area and one accepts that
position if \mbox{$x\le\frac{1+\kappa(\vec{\theta})}{1+\kappa_{\rm
max}}$}, where $x$ is a random number between $x\in[0,1]$ and
$\kappa_{\rm max}$ the maximum value within the $\kappa$ field.
Following this procedure one gets mock data for which
$\kappa=\kappa_{\rm g}$ and therefore $\ave{{\cal N}^nM_{\rm
ap}^m}=\ave{{\cal N}^{n+m}}$. In particular we must get, apart from
the statistical noise due to finite galaxy numbers, $\ave{{\cal
N}^2(\theta_1)M_{\rm ap}(\theta_2)}=\ave{{\cal N}(\theta_1)M^2_{\rm
ap}(\theta_2)}$ when running our codes with the mock data.

Parallelly, we smooth the test shear field within apertures according
to the definitions \Ref{mapinshear} with our aperture filter and
estimate the test data aperture statistics directly by
cross-correlating the smoothed fields. This also has to be comparable
(apart from shot noise) to our code output.  The result of this test
can be found in Fig. \ref{kappatest}.

As a further test we take the same mock data but rotate the
ellipticities of the sources by $45$ degrees, i.e. we multiply the
complex ellipticities by the phase factor \mbox{$-\e^{-2\i\phi}$} with
\mbox{$\phi=45^\circ$}. This generates a purely B-mode signal that
should only be picked up by the B-mode channels of the aperture
statistics, yielding a plot similar to Fig. \ref{kappatest}. The
parity mode in \mbox{$\ave{{\cal N}M^2_{\rm ap}}$} has to be
unaffected. This is indeed the case (figure not shown).

The test results make us confident that the computer code is working
and that we achieve a good accuracy even though we are forced to bin
the correlation functions and to use a tree-code that necessarily
makes some additional approximations.

\section{Results and discussion}
\label{resultsection}

We applied the previously outlined method to the RCS shear and lens
catalogues. 

Lenses were selected between photometric redshifts \mbox{$0<z<0.4$},
whereas sources were from the range \mbox{$0.5<z<1.4$}. Compared to
\citet{2005ApJ...635...73H}, in which photometric redshifts smaller
than $0.2$ were excluded, we were less strict about the lowest
redshift of the lenses. This is likely to have introduced some
misidentified lenses into our sample (less than $10\%$, see
Fig. \ref{pofz}) as RCS is lacking a U-band filter. Moreover,
including sources with photometric redshifts larger than $z\sim1.0$ is
also rather optimistic because \changed{photometric redshifts} within
that range can become quite unreliable as well. Therefore the tail of
the redshift distribution in Fig. \ref{pofz} may be slightly
inaccurate. Still, sources with photo-z's greater than one are likely
to be high-redshift galaxies.

However, for our purpose, namely demonstrating a robust detection of
GGGL, the biases in the redshift distribution of lenses and sources
are acceptable. These biases in the estimated redshift distribution
only become an issue if one wants to thoroughly model the GGGL-signal.

\begin{figure*}
  \begin{center}
    \epsfig{file=8197fig5.ps,width=65mm,angle=-90}
    \hspace{.1cm}
    \epsfig{file=8197fig6.ps,width=65mm,angle=-90}
  \end{center}
  \caption{\label{resultfig} \emph{Left}: Aperture statistics
    \mbox{$\ave{{\cal N}^2M_{\rm ap}}(\theta,\theta,\theta)$} for
    different aperture radii $\theta$ as measured in RCS. The upper
    panel is the E-mode, the lower panel is the parity mode which is
    consistent with zero. Error bars denote the field-to-field
    variance between the ten RCS fields. Statistical errors are
    strongly correlated. The lines are tentative halo model-based
    predictions with arbitrary HODs for a $\Lambda\rm CDM$
    cosmological model (see text). \emph{Right}: Aperture statistics
    \mbox{$\ave{{\cal N}M_{\rm ap}^2}(\theta,\theta,\theta)$} for
    different aperture radii $\theta$ as measured in RCS. The upper
    panel contains the E-mode measurement, while B-mode (stars) and
    parity mode (squares) are plotted inside the lower panel. Error
    bars that extend to the bottom of the upper panel denote data
    points that are consistent with zero.}
\end{figure*}

\subsection{Aperture statistics}

\begin{figure}
  \begin{center}
    \includegraphics[width=60mm, angle=-90]{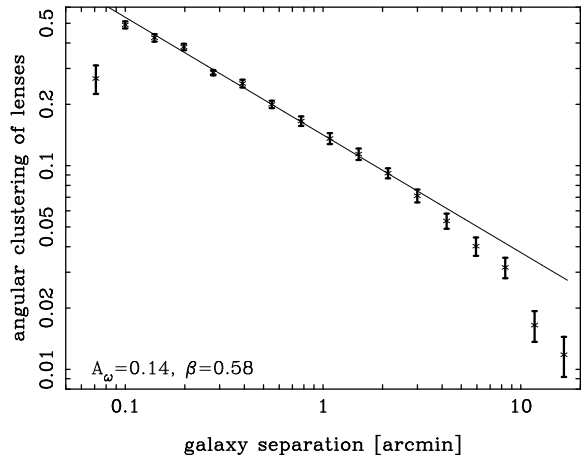}
  \end{center}
  \caption{\label{lensclustering} Combined measurement of angular
  clustering of our sample of lenses (no correction for the integral
  constraint). Error bars were obtained by looking at the
  field-to-field variance. The solid line is a power-law fit,
  \mbox{$\omega(\theta)=A_\omega\theta^{-\beta}$}, to the regime
  \mbox{$\theta\in[0^\prime\!.1,3^\prime]$}.}
\end{figure}

\changed{As a first result} we would like to draw the reader's
attention to the angular clustering of lenses which is plotted in Fig.
\ref{lensclustering}. This measurement was required for the estimator
$\tilde{\cal G}$ in Eq. \Ref{gestimator}. As widely accepted, the
angular correlation function $\omega(\theta)$ is, for the separations
we are considering here, well approximated by a simple power-law,
depending on galaxy type, colour and luminosity
\citep[e.g.][]{mad03}. As can be seen in Fig. \ref{lensclustering},
the power-law behaviour is also found for our lens galaxy sample. The
angular clustering plotted is still affected by the so-called integral
constraint \citep{gp77}, which shifts the estimate of $\omega$
downwards by a constant value depending on the geometry and size of
the fields. \changed{For small \mbox{$\theta\lesssim3^\prime$} this
bias is negligible so that we used only the regime
\mbox{$\theta\in[0^\prime\!\!.1,3^\prime]$} to find the maximum likelihood
parameters of the power-law.}

\changed{For $\tilde{\cal G}$ this power-law fit was used. Possible
deviations of the true clustering from a power-law for
$\theta\ge2^\prime$ were negligible because for the estimator one
actually needs $1+\omega$ instead of $\omega$. Since $\omega$ is
roughly smaller than $\sim0.05$ and decreasing for
$\theta\ge2^\prime$, we gather that a certain remaining inaccuracy in
$\omega$ has no big impact on $1+\omega$.}  The power-law index is,
with \mbox{$\beta=0.58$}, fairly shallow, which is typical for a
relatively blue sample of galaxies \citep[e.g.][]{mad03}.

In a second step, the correlation functions $\tilde{\cal G}$ and
$\tilde{G}_\pm$ were computed separately for each of the ten RCS
fields. The total combined signal was computed by taking the average
of all fields, each bin weighted by the number of triangles it
contained. For the binning we used a range of
\mbox{$0^{\prime\prime}\!.8\leq\vartheta\leq54^\prime$} with $100$
bins, thus overall $10^6$ triangle configurations. By repeatedly
drawing ten fields at random from the ten available, i.e. with
replacement, and combining their signal we obtained a bootstrap sample
of measurements. The variance among the bootstrapped signals was used
to estimate the sum of cosmic variance and shot noise, thus the
remaining statistical uncertainty of the correlation functions.

\begin{figure*}
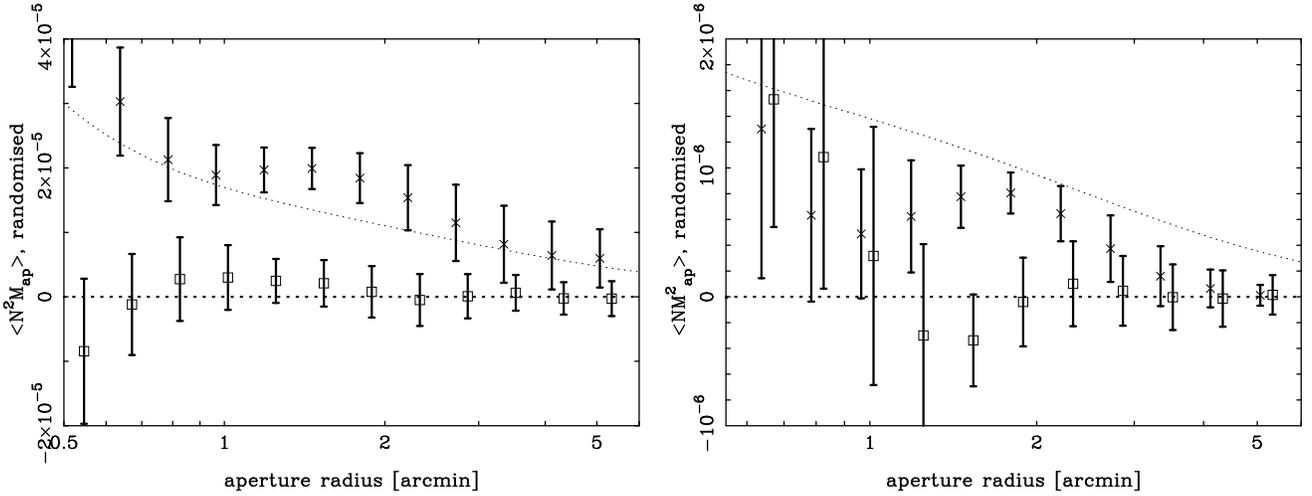

  \begin{center}
    \epsfig{file=8197fig8.ps,width=65mm,angle=-90}
    \hspace{.1cm}
    \epsfig{file=8197fig9.ps,width=65mm,angle=-90}
  \end{center}
  \caption{\label{randomised} Residual signal (squares) of GGGL in RCS
  when the ellipticities of the sources are randomised (\emph{left}:
  $\ave{{\cal N}^2M_{\rm ap}}$, \emph{right}: $\ave{{\cal N}M^2_{\rm
  ap}}$). For comparison, the original signal before randomisation is
  also plotted (crosses). The line is a crude halo-model prediction of
  a blue galaxy population as in Fig. \ref{resultfig}. The error bars
  of the randomised signal quantify the background noise of a
  null-signal. This indicates that we have a significant detection of
  \mbox{$\ave{{\cal N}^2M_{\rm ap}}$} in the left panel but only a
  weak detection of \mbox{$\ave{{\cal N}M_{\rm ap}^2}$}, most
  significant at about $2^\prime$, in the right panel.}
\end{figure*}

Finally, the correlation functions were transformed to the aperture
statistics considering only equally sized apertures,
i.e. \mbox{$\theta_1=\theta_2=\theta_3$}, see
Fig. \ref{resultfig}. For the scope of this work, equally sized
apertures are absolutely sufficient. In future work, however, one
would like to harvest the full information that is contained in these
statistics by exploring different $\theta_i$ which then would cover
the full (projected) bispectrum.

For a start, we would like to focus on $\tilde{\cal G}$.  The left
panel in Fig. \ref{resultfig} reveals a clean detection of
\mbox{$\ave{{\cal N}^2M_{\rm ap}}$} for aperture radii between
\mbox{$0^\prime\!\!.5\lesssim\theta\lesssim5^\prime$} (with the
adopted filter this corresponds to typical angular scales between
$1^\prime$ and $11^\prime$) demonstrating the presence of pure \thrd
correlations between shear and lens distribution in RCS. The parity
mode of this statistic is consistent with the zero as expected.
Fig. \ref{resultfig} is one of the central results of this paper.

We would like to further support that this is a real,
i.e. cosmological, signal by comparing the measurement to crude halo
model-based predictions \citep[see][ for a review]{cooraysheth02}. The
halo model was used to predict a spatial cross-correlation bispectrum,
$B_{{\rm gg}\delta}$ \citep[Eq. 12 in][]{gggl05}, for a particular
fiducial cosmological model and halo occupation distribution (HOD) of
galaxies \citep[see][]{bw02}. By applying Eqs. (21), (52) in
\citet{gggl05}, $B_{{\rm gg}\delta}$ was transformed, taking into
account the correct redshift distribution of lenses and sources (Fig.
\ref{pofz}), to yield the aperture statistics. A standard concordance
$\Lambda\rm CDM$ model was employed \citep{bbks86} with parameters
\mbox{$\Omega_\Lambda=0.7$}, for the dark energy density,
\mbox{$\Omega_{\rm m}=0.30$}, for the (cold) dark matter density,
\mbox{$\sigma_8=0.9$} for the power spectrum normalisation, and
\mbox{$\Gamma=0.21$} for the shape parameter. This is in agreement
with constraints based on the first WMAP release
\citep{2003ApJS..148..175S}.

The latest constraints favour a somewhat smaller value for $\sigma_8$
\citep{2007astro.ph..3570B, hetter06} which would shift the expected
amplitude of GGGL towards smaller values. If we apply the scaling
relation of \citet{1997ApJ...484..560J}, given for the convergence
bispectrum, as a rough estimate of this shift, \mbox{$B_\kappa\propto
\sigma_8^{5.9}$}, we obtain a correction factor of about two for
\mbox{$\sigma_8=0.8$} ($\cal N$ and $M_{\rm ap}$ should have the same
$\sigma_8$-dependence for unbiased galaxies).

The halo-model predictions depend strongly on the adopted HOD. The
basic set up for this model was that outlined in
\cite{2003MNRAS.340..580T}, which splits the occupation function,
$N(M)$, into contributions from ``red'', $N_R$, and ``blue'',$N_B$,
galaxies:
\begin{eqnarray}\nonumber
\ave{N_B}(M)&=&\left(\frac{m}{m_B}\right)^{\gamma_B}+
{\rm A}\exp{\left(-{\rm A}_0(\log_{10}(m)-m_{B_s})^2\right)}\\
\ave{N_R}(M)&=&\left(\frac{m}{m_R}\right)^{\gamma_R}
\exp{\left(-\left[\frac{m_{R_0}}{m}\right]^{1,2}\right)}\;.
\end{eqnarray}

As parameters we used $m_B=2.34\times10^{13}\,\msol$, ${\rm A}=0.65$,
${\rm A}_0=6.6$, $m_{B_s}=11.73$, $m_R=1.8\times10^{13}\,\msol$ and
$m_{R_0}=4.9\times10^{12}\,\msol$.  Blue galaxies have a peak halo
occupancy of around $10^{12}\,\msol$ and a shallow power law
($\gamma_B=0.93$) at high halo masses. In this simple prescription,
red galaxies are relatively more numerous in higher mass halos
($\gamma_R=1.1$) and are excluded from low mass halos by an
exponential cutoff around $5\times10^{12}\,\msol$. Factorial moments
of the occupation distribution - the cross bisprectra $B_{gg\delta}$
and $B_{\delta\delta g}$ require the mean and variance - were as
prescribed in the model of \cite{2001ApJ...546...20S}.  In this way,
the moments are Poissonian for higher mass halos, becoming
sub-Poissonian for masses below $10^{13}\,\msol$, i.e.
\mbox{$\ave{N^2}(M)=\alpha^2 [\ave{N}(M)]^2$}, where
\mbox{$\alpha=0.5\log_{10}{(m/10^{11}\,\msol)}$}.

We stress at this point that we made no attempt to ``fit'' parameters
to the data, we merely intended to bracket a range of possible
results. To choose a range of plausible scenarios, we constructed the
\changed{theoretical aperture statistics for ``red'' galaxies,
``blue'' galaxies and for ``all'' galaxies} (in which the occupation
functions for red and blue galaxies are added together directly). We
also showed predictions for the unbiased case, in which the occupation
function \mbox{$N(M)\propto M$} with Poisson moments for
$\ave{N^2(M)}$. Galaxies were assumed to follow the CDM halo density
profile (NFW) with no assumption of a central galaxy. Other parameters
that define the halo model set up (e.g. concentration of the NFW
profile) were as used in \cite{2003MNRAS.340..580T}.

Our measurement of $\ave{{\cal N}^2M_{\rm ap}}$ lies somewhat above
the lower bound of the expected physical range of values, giving
support as to the cosmological origin of the signal. Moreover, taken
at face value, our result appears to fit the picture that the lens
population consists of rather blue galaxies as has been concluded from
the shallow slope of the angular correlation function $\omega$.

We randomised the ellipticities of the sources and repeated the
analysis. Since the coherent pattern, and its correlation to the lens
distribution, is responsible for the signal, destroying the coherence
by randomising the ellipticity phase should diminish the signal. That
this is the case can be seen in Fig. \ref{randomised} (left panel).

Analogous to $\ave{{\cal N}^2M_{\rm ap}}$ we computed and predicted
$\ave{{\cal N}M_{\rm ap}^2}$, the result for which is shown in
Fig. \ref{resultfig} (right panel). Here a signal significantly
different from zero was only found for aperture radii
\mbox{$1^\prime\le\theta \le3^\prime$} and at about
$\theta\sim0^\prime\!\!.5$. Below $\theta\sim1^\prime\!\!.5$ the
parity mode is not fully consistent with zero. Hence, we may have a
non-negligible contamination by systematics in the PSF correction
and/or intrinsic alignments of the sources that may hamper a clean
detection. For radii where we find a non-zero signal, the signal is on
average smaller than the lowest theoretical value from our crude
models. However, as discussed above, a lower $\sigma_8$ easily brings
the model down towards smaller values.  The signal disappeared if the
ellipticities of the sources were randomised (Fig. \ref{randomised},
right panel). Therefore, we found a tentative detection of $\ave{{\cal
N}M_{\rm ap}^2}$ in our data.

\begin{figure}
  \begin{center}
    \epsfig{file=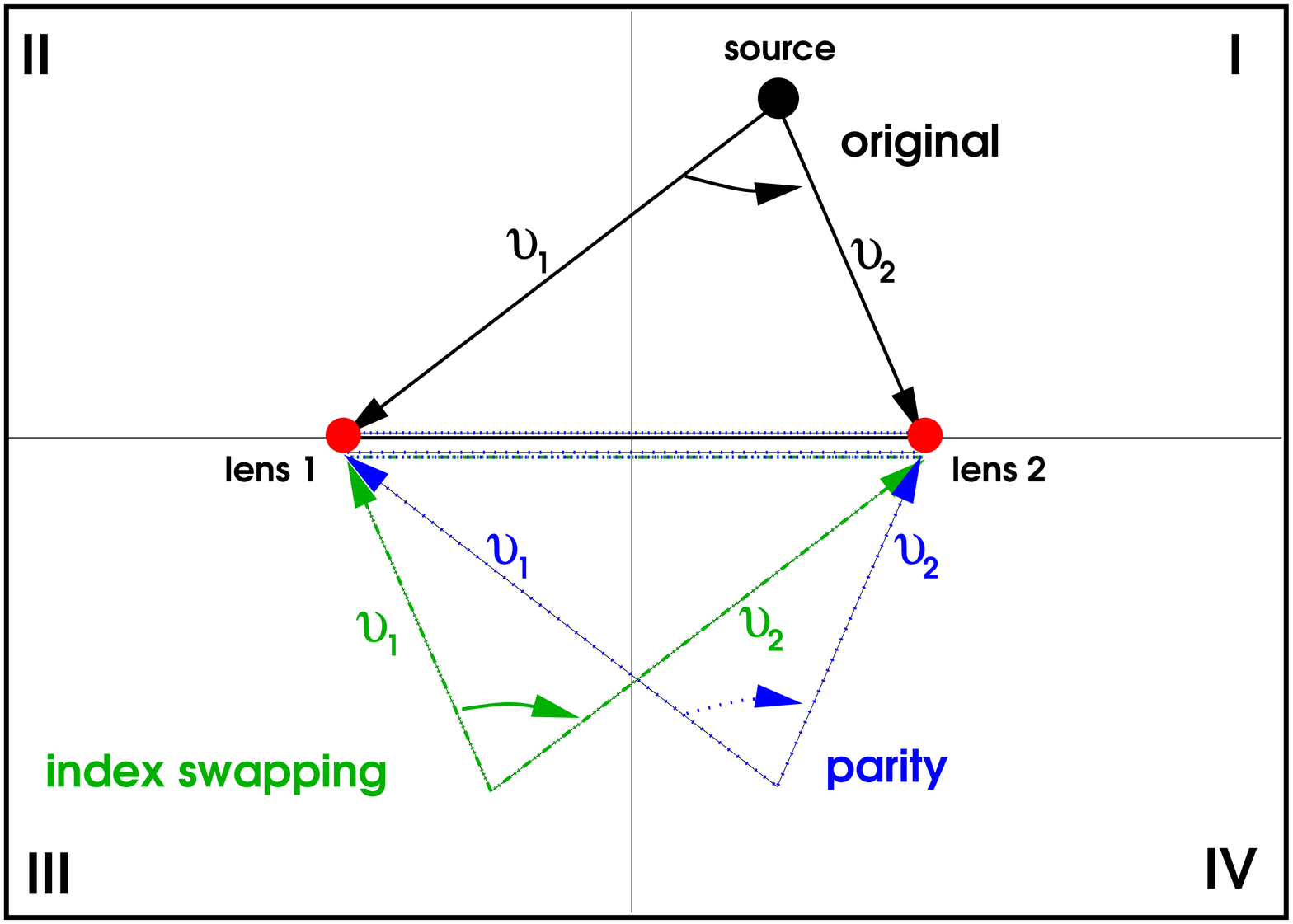,width=80mm,angle=0}
  \end{center}
  \caption{\label{Gplotexplain} Sketch illustrating how $\tilde{\cal
    G}$ or $\cal G$ are plotted. See text for details.}
\end{figure}

\begin{figure}
  \begin{center}
    \epsfig{file=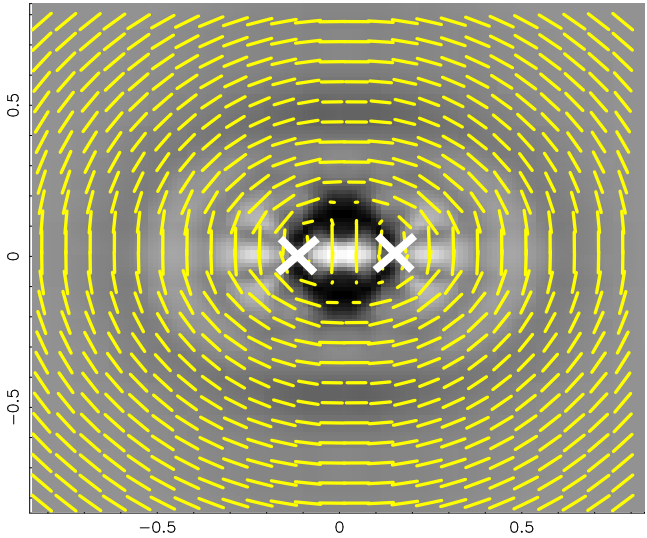,width=85mm,angle=0}
  \end{center}
  \caption{\label{Gplot2ndorder} Signal in $\tilde{\cal G}$
    originating from pure $2^{\rm nd}$-order statistics (GGL) that was
    subtracted from $\tilde{\cal G}$ to obtain Fig. \ref{Gplot}
    (left). The gray-scale intensity is the modulus of $\tilde{\cal
    G}$, the sticks indicate the average shear at the source position
    in the lens-lens-source triangle configuration. The units of the
    axis are in \mbox{$h^{-1}\rm Mpc$} which corresponds to the mean
    physical scale at the lens distance of about \mbox{$z=0.30$}. The
    two lenses are located at the positions of the crosses, left and
    right from the centre.}
\end{figure}

\subsection{Mapping the excess matter distribution about two
  lenses}

The aperture statistics clearly have advantages: the B- and
parity-modes allow a check for remaining systematics in the data, and
\snd statistics do not make any contributions so that we can be sure
to pick up a signal solely from connected \thrd terms. This is what we
did in the forgoing subsection. The result suggests that we have a
significant detection of $\cal G$.

The disadvantage of using aperture statistics is, however, that they
are hard to visualise in terms of a typical (projected) matter
distribution (lensing convergence) about two lenses, say. Therefore,
we introduce here an alternative way of depicting $\cal G$ which is
similar to the work that has been proposed by
\cite{2006MNRAS.367.1222J}. 

A similar way of visualising $G_\pm$ \changed{probably could} be
thought up as well. However, since we found only a weak detection of
GGGL with two sources and one lens we postpone this task to a future
paper and focus here on $\cal G$ alone.

\begin{figure*}
  \begin{center}
    \epsfig{file=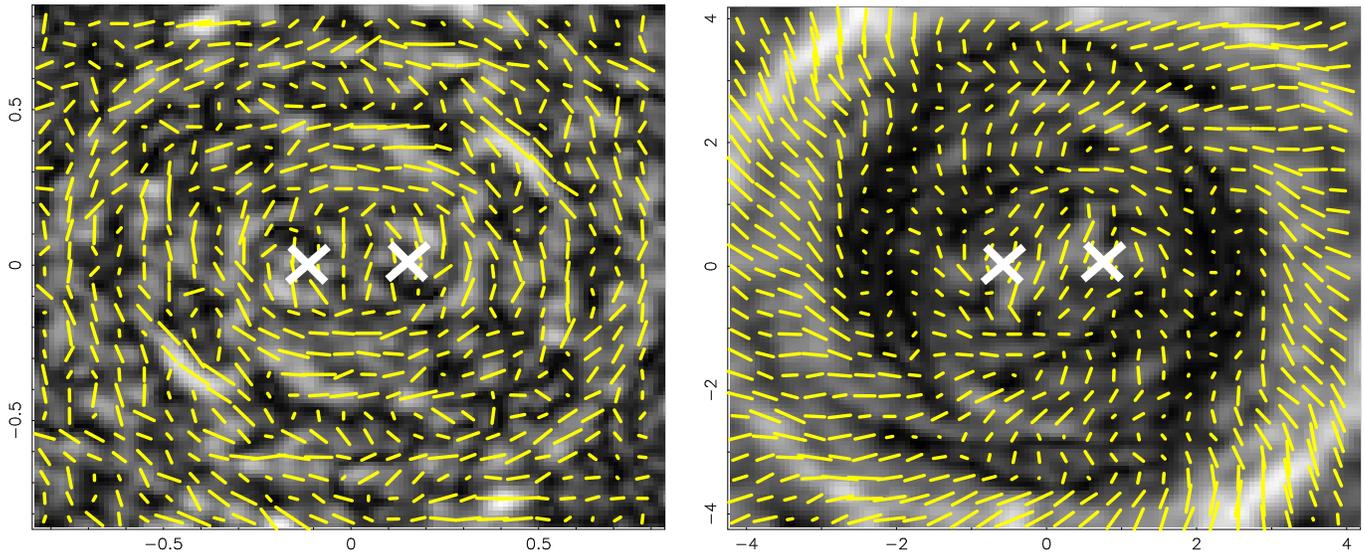,width=180mm,angle=0}
  \end{center}
  \caption{\label{Gplot} Plots of $\cal G$ after subtraction of the
  $2^{\rm nd}$-order signal from $\tilde{\cal G}$. The units used are
  \mbox{$h^{-1}\rm Mpc$}, which corresponds to the mean comoving
  physical distance at the lenses' distance of, on average,
  \mbox{$z=0.30$}. \emph{Left:} Lenses were selected to have a mutual
  angular separation between $40^\pprime$ and $80^\pprime$
  corresponding a projected physical scale of about $250\,h^{-1}\rm
  kpc$. \emph{Right:} Lenses were chosen to have a separation between
  $4^\prime$ and $8^\prime$, or equivalently a projected comoving
  separation between $1-2\,h^{-1}\rm Mpc$.}
\end{figure*}

The following summarises what essentially is done if we estimate
$\tilde{\cal G}$ from the data for fixed lens-lens separations.  We
pick out only lens-lens-source triangles from our data set in which
the lenses have a fixed separation or a separation from a small
range. Each triangle is placed inside the plot such that the line
connecting the lenses is parallel to the $x$-axis and that the centre
of this line coincides with the centre of the plot, as seen for the
triangles in Fig. \ref{Gplotexplain}. The ellipticities of the sources
of all triangles are then multiplied by
\mbox{$1+\omega(|\theta_2-\theta_1|)$} (rescaled according to
Eq. \ref{rescale}) and (weighted) averaged at the source
positions. For this paper, we used \mbox{$128\times128$} grid cells
for binning the ellipticities. Following this procedure we effectively
stacked all shear patterns about a lens-lens configuration -- rotated
appropriately -- to obtain an average shear field about two
lenses. This is, in essence, the meaning of $\tilde{\cal{G}}$. The
full $\tilde{\cal G}$ is a bundle of such plots with continuously
changing lens-lens separations.

\begin{figure*}
  \begin{center}
    \epsfig{file=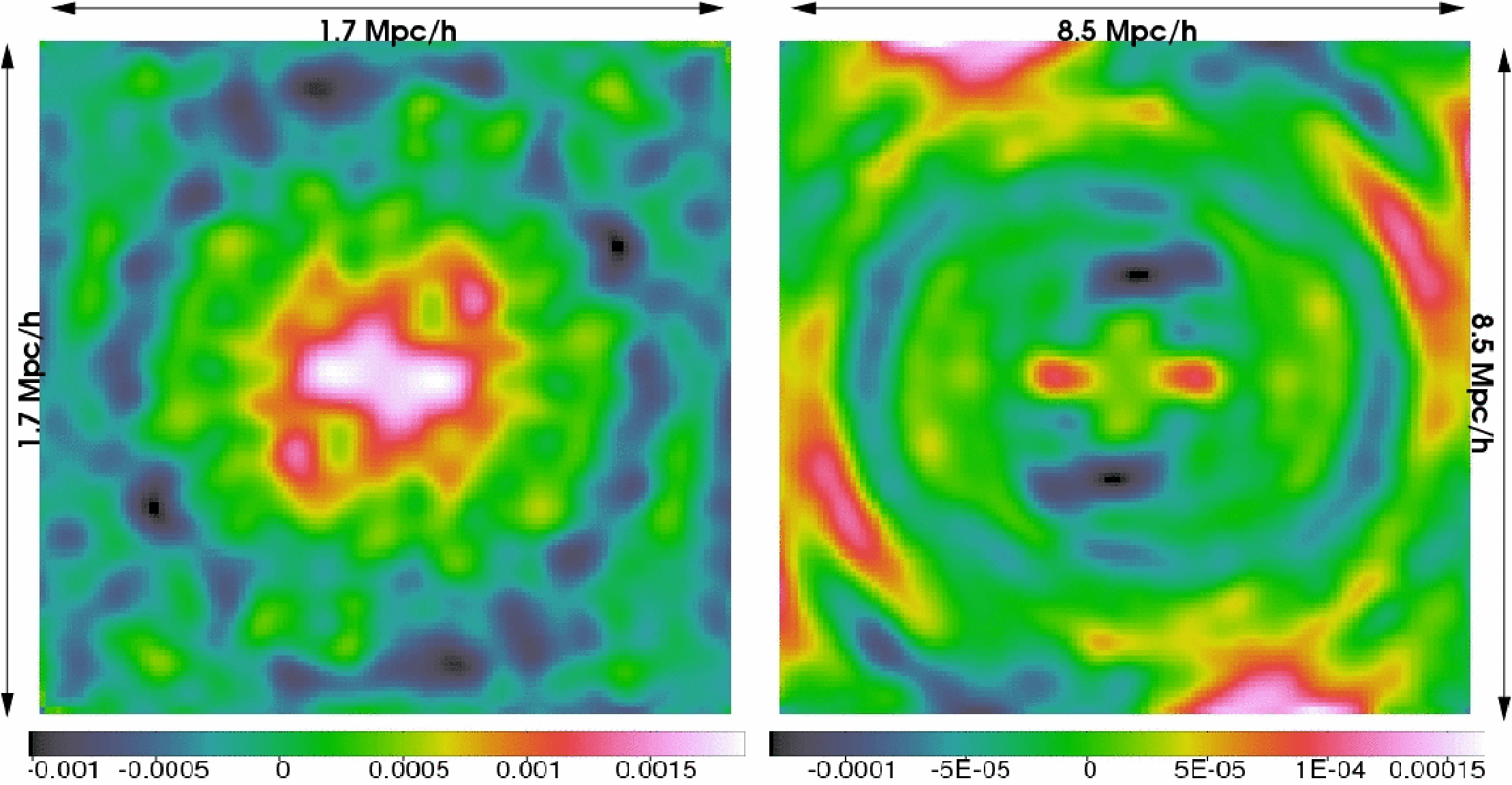,width=175mm,angle=0}
  \end{center}
  \caption{\label{kappaplot} Convergence fields obtained by
  transforming the shear fields in Fig. \ref{Gplot}. They are related
  to the (average) excess in matter density around two galaxies of
  fixed \emph{angular} separation after subtraction of the matter
  density profile that is observed about individual
  galaxies. \emph{Left:} Residual convergence for two lenses with
  projected comoving distance of roughly \mbox{$250\,h^{-1}\rm
  kpc$}. The box-size is $1.7\,h^{-1}{\rm Mpc}\times1.7\,h^{-1}{\rm
  Mpc}$. \emph{Right:} Residual convergence at about
  \mbox{$1.5\,h^{-1}\rm Mpc$} projected lens-lens distance. The
  box-size is $8.5\,h^{-1}{\rm Mpc}\times8.5\,h^{-1}{\rm Mpc}$. Note
  that the convergence in this figure is lower by roughly an order of
  magnitude compared to the left figure.}
\end{figure*}

Note that the ellipticity at the source position, stored in $\cal G$,
is rotated by $\phi_3/2$ (Fig. \ref{corrsketch}, right panel). For the
following plots, on the other hand, we used the shear in Cartesian
coordinates relative to axis defined by the lens positions, as in
\citet{2006MNRAS.367.1222J}. Therefore, when generating the plot we
were rotating our measurements for $\tilde{\cal G}$ appropriately.

The resulting plot has symmetries. Firstly, we do not distinguish
between ''lens 1'' and ``lens 2''. Both lenses are drawn from the
\emph{same galaxy sample}. This means for every triangle, we will find
the same triangle but with the positions of ``lens 1'' and ``lens 2''
exchanged. Therefore, the two lenses and the source of the triangle
named ``original'' in Fig. \ref{Gplotexplain} will make the same
contribution but complex conjugated at the source position of the
triangle named ``index swapping''. Thus, quadrants I and III will be
identical apart from a complex conjugate and mirroring the positions
about the $x$- and $y$-axis. The same holds for quadrants II and
IV. This would no longer be true, of course, if we chose the two
lenses from different catalogues in order to, for instance, study the
matter distribution around a blue and a red galaxy.

A second symmetry can be observed if the Universe (or the
PSF-corrected shear catalogue) is parity invariant. Mirroring the
triangle ``original'' with respect to the line connecting the two
lenses ($x$-axis) results in another triangle coined ``parity''. For
parity invariance being true the ellipticity at the source position of
``parity'' is \emph{on average} identical to the ellipticity at the
source position of triangle ``original''. In this case, quadrant IV is
statistically consistent with quadrant I and quadrant II with quadrant
III (after mirroring about the $x$-axis). Taking parity symmetry for
granted could be used to increase the signal-to-noise in the plots by
taking the mean of quadrants IV and I (or II and III).

Since the way of binning in the plot is completely different from the
way used to get the aperture statistics out of RCS, we made two reruns
of the estimation of $\tilde{\cal G}$ with our data. For the first run
we only considered lens-lens separations between $40^\pprime$ and
$80^\pprime$, the second run selected triangles in which the lenses
had a separation between $4^\prime$ and $8^\prime$. For a mean lens
redshift of \mbox{$z\sim0.3$} this corresponds to a projected physical
comoving separation of roughly $250\,h^{-1}\rm kpc$ and
$1.5\,h^{-1}\rm Mpc$, respectively. As usual, the results from the ten
individual fields were averaged by weighting with the number of
triangles inside each bin and the statistical weights of the sources.

Since we effectively stacked the shear fields about all pairs of
lenses, aligned along the lens-lens axis, we obtained the average
shear about two lenses. The shear pattern still contained a
contribution stemming from GGL alone. This contribution could,
however, easily be subtracted according to Eq. \Ref{gtilde} after
estimating the mean tangential shear, $\ave{\gamma_{\rm
t}}(\vartheta)$, about single lens galaxies \citep[see
e.g.][]{simon07}. A typical shear pattern due to \snd GGL can be seen
in Fig. \ref{Gplot2ndorder}.  This is the shear pattern that is to be
expected if the average shear about two lenses is just the sum of two
mean shear patterns about individual lenses. They contain all
contributions that are statistically independent of the presence of
the other lens. Therefore, contributions (contaminations) to the shear
from lens pairs that are just accidentally close to each other by
projection effects, but actually too separated in space to be
physically connected, are removed.

Now, Fig. \ref{Gplot} shows the shear patterns after removing this
signal. Clearly, there is a residual coherent pattern which is most
pronounced for the smaller lens-lens separations. This proves that one
finds an additional shear signal around two galaxies if they get close
to each other. Hence, the average gravitational potential about two
close lenses is not just the sum of two average potentials about
individual lenses.

Unfortunately, \emph{all} physically close galaxies with a fixed
projected angular separation contribute to the excess shear--
independent of whether they are in galaxy groups or
clusters. Exploiting lens redshifts and rejecting lenses from regions
of high number densities on the sky might help to focus on galaxy
groups, for example. This, however, is beyond the scope of this paper.

\begin{figure*}
  \begin{center}
    \epsfig{file=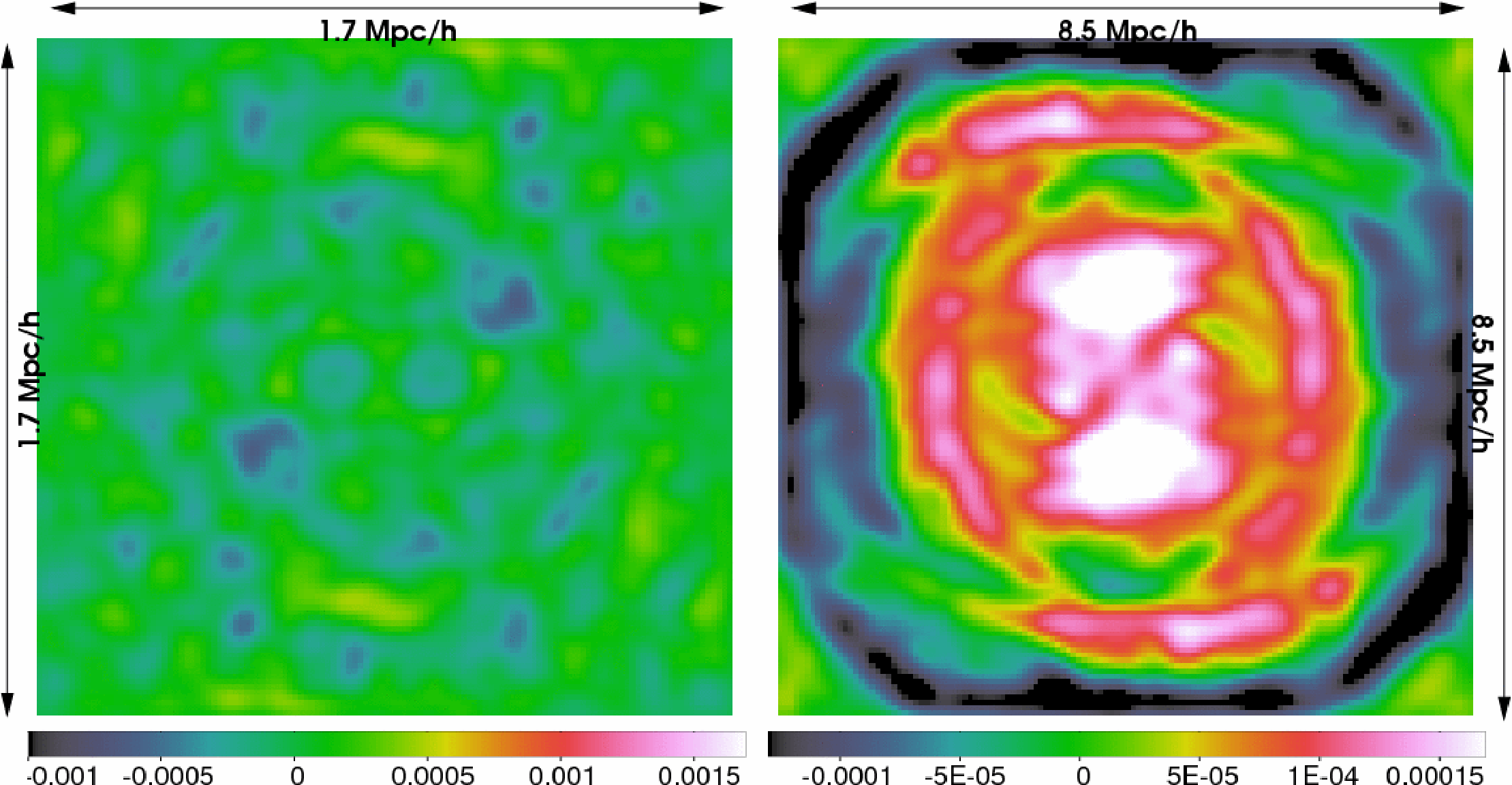,width=175mm,angle=0}
  \end{center}
  \caption{\label{kappaplotbmode} Plots similar to the plots in
  Fig. \ref{kappaplot} except the the shear has been rotated by
  $45^\circ$ (B-mode) before transforming to the convergence
  fields. The thereby obtained convergence quantifies the statistical
  noise in the plots of Fig. \ref{kappaplot}.}
\end{figure*}

One can relate the residual shear pattern in Fig. \ref{Gplot} to an
excess in projected convergence (matter density) using the well known
relation between convergence and cosmic shear in weak gravitational
lensing \citep{bs01,1993ApJ...404..441K}:
\begin{equation}
  \gamma_\ell=\frac{\ell_1^2-\ell_2^2+2{\rm
      i}\ell_1\ell_2}{\ell_1^2+\ell_2^2}~\kappa_\ell\;,
\end{equation}
where $\gamma_\ell$ and $\kappa_\ell$ are the Fourier coefficients of
the shear and convergence fields, respectively, on a grid and
\mbox{$\ell=(\ell_1,\ell_2)$} is a particular angular mode of the grid
in Cartesian coordinates. We obtained the $\gamma_\ell$'s by employing
Fast-Fourier-Transforms (and zero-padding to reduce undesired edge
effects) after binning the residual shear patterns onto a
\mbox{$512\times512$} grid. We assumed that the convergence is zero
averaged over the box area which makes \mbox{$\kappa_\ell=0$} for
\mbox{$\ell=0$}.  Fig. \ref{kappaplot} shows the thereby computed
maps. The plots were smoothed with a kernel of a size of a few pixels.

\changed{As a cross-check} we also transformed the shear pattern
produced by the \snd terms in $\tilde{\cal G}$
(Fig. \ref{Gplot2ndorder}) and found, as expected, that the
corresponding convergence fields were just two identical radially
symmetric ``matter haloes'' placed at the lens positions in the plot.

In the same way as in the previously discussed shear plots, parity
invariance can also be checked in the convergence plots: quadrants I
and IV (or II and III), mirrored about the $x$-axis, have to be
statistically consistent. If we would like to enforce parity
invariance, we could take the average of the two quadrants. Secondly,
if one obtains the convergence field from the shear field via a
Fourier transformation as described before, the convergence field will
be a field of complex numbers. In the absence of any B-modes, however,
the imaginary part will be zero or pure noise. Thus, the imaginary
part of the convergence can be used to either check for residual
B-modes or to estimate the noise level of the E-mode (real part).

This was done for Fig. \ref{kappaplotbmode}. We found that the
residual convergence for the small lens-lens separation is highly
significant within the central region of Fig. \ref{kappaplot}, left
panel, whereas the convergence in the right panel of
Fig. \ref{kappaplot} is noise dominated. This means we did not find
any excess convergence beyond the noise level for the lens-lens pairs
of large separation.

To sum up, one can see that closer lens pairs are embedded inside a
common halo of excess matter, while the lenses with larger separation
appear relatively disconnected; the convergence for the lenses of
larger separation is lower by at least one order of magnitude and
slips below noise level in our measurement. This result definitely
deserves further investigation which we will do in a forthcoming
paper.

\section{Conclusions}

We found a significant signal of GGGL in RCS -- at least for the case
for which we considered two lenses and one source. The signal is of an
order of magnitude which is expected from a crude halo model-based
prescription. This suggests a cosmological origin of the observed
correlation.

In particular, our finding demonstrates that wide-field surveys of at
least the size of RCS allow us to exploit GGGL.  As can be seen in
Fig. \ref{resultfig} (left), the remaining statistical uncertainties
of the measurement are much smaller than the shift of the signal
expected for different HODs of the adopted halo model. This means with
GGGL we now have a new tool to strongly constrain galaxy HODs, and
possibly even spatial distributions of galaxies inside haloes in
general, which is a parameter in the framework of the halo model. As
the wide-field shear surveys of the next generation will be
substantially larger than RCS those constraints will become
tighter. Further subdivisions of lens samples into different galaxy
types and redshifts will therefore still give a reasonable
signal-to-noise ratio.

Leaving the interpretation in the context of the halo model aside, the
measurement of GGGL can be translated into a map of excess convergence
around two galaxies of a certain mutual (projected) distance. For RCS,
we demonstrated that there is a significant excess in convergence
about two lenses if galaxies are as close as roughly
\mbox{$250\,h^{-1}\rm kpc$}. Although the details need still to be
worked out, this promises \changed{to be a novel} way of studying the
matter environment of groups of galaxies.
 
\begin{acknowledgements}
  We would like to thank Jan Hartlap for providing us with simulated
  shear catalogues used as mock data. We are also grateful to Emilio
  Pastor Mira, who kindly computed the aperture statistics in our mock
  data using his aperture based code, Oliver Cordes who helped us with
  the Linux cluster and Lindsay King for comments on the paper. This
  work was supported by the Deutsche Forschungsgemeinschaft (DFG)
  under the project SCHN 342/6--1 and by the Priority Programme SPP
  1177 `Galaxy evolution' of the Deutsche Forschungsgemeinschaft under
  the project SCHN 342/7--1.  Patrick Simon was also supported by
  PPARC. Henk Hoekstra acknowledges support from NSERC and CIAR.
\end{acknowledgements}
\bibliographystyle{aa}
\bibliography{8197gggl}
\end{document}